\documentclass[pre,aps,twocolumn,showpacs,floatfix,10pt,superscriptaddress]{revtex4-2}
\usepackage{graphicx}
\usepackage{color}
\usepackage{amsmath, amsthm, amssymb}
\usepackage{float}
\usepackage{comment}
\usepackage{hyperref}

\usepackage{enumerate}

\newcommand{\be}{\begin{equation}}
\newcommand{\ee}{\end{equation}}
\newcommand{\bea}{\begin{eqnarray}}
\newcommand{\eea}{\end{eqnarray}}

\begin{document}

\title{Probabilities of rare events in product kernel aggregation: An exact formula and phase diagram}

\author{R. Goutham}
\email{gouthamr@imsc.res.in}
\affiliation{The Institute of Mathematical Sciences, C.I.T. Campus,
	Taramani, Chennai 600113, India}
\affiliation{Homi Bhabha National Institute, Training School Complex, Anushakti Nagar, Mumbai 400094, India}

\author{R. Rajesh}
\email{rrajesh@imsc.res.in}
\affiliation{The Institute of Mathematical Sciences, C.I.T. Campus,
	Taramani, Chennai 600113, India}
\affiliation{Homi Bhabha National Institute, Training School Complex, Anushakti Nagar, Mumbai 400094, India}

\author{V. Subashri}
\email{subashriv@tifrh.res.in}
\affiliation{Tata Institute of Fundamental Research, Hyderabad 500046, India}

\author{Oleg Zaboronski}
\email{olegz@maths.warwick.ac.uk}
\affiliation{Mathematics Institute, University of Warwick, Gibbet Hill Road, Coventry CV4 7AL, United Kingdom}

\date{\today}

\begin{abstract}

We present an exact method for calculating the large deviation function describing rare fluctuations in the number of particles for product-kernel aggregation. Starting from the master equation, we derive an exact integral representation for the probability $P(M,N,t)$ of observing $N$ particles at time $t$ starting from $M$ monomers for any  finite $M, N, t$. From this, we obtain an exact expression for the exponential moment $\langle p^N\rangle$ for integer $p$. Employing a replica conjecture---numerically validated by finite-$M$ scaling---we extend this result to real $p \geq 0$. The convex envelope of the large deviation function, obtained via a Legendre-Fenchel transform of the exponential moment, shows singular behavior. The singular structure allows us to construct the full phase diagram of product-kernel aggregation, which contains a tricritical point, separating continuous and discontinuous transitions.  We also compute the asymptotic form of the LDF for small $N/M$.
\end{abstract}


\maketitle

\section{\label{sec:intro}Introduction}

There are a wide variety of physical phenomena in which the dominant dynamical process is aggregation of constituent particles, a non-equilibrium, irreversible process. Examples include blood coagulation~\cite{guria2009mathematical}, cloud formation~\cite{falkovich2002acceleration,pruppacher2012microphysics}, aerosol dynamics~\cite{williams1988unified}, the coagulation of dust and gas in the formation of Saturn’s rings~\cite{brilliantov2015size,connaughton2018stationary}, the aggregation of particulate matter in oceans~\cite{burd2009particle}, protein aggregation~\cite{benjwal2006monitoring,wang2015following}, charged biopolymers~\cite{tom2016aggregation,tom2017aggregation}, and ductile fracture~\cite{pineau2016failure}.  A minimal model that isolates the effect of aggregation is the cluster–cluster aggregation (CCA) wherein particles coalesce upon contact to form larger clusters, the study of which dates back to the pioneering work of Smoluchowski in 1917~\cite{smoluchowski1917mathematical}.  CCA also finds applications in diverse applied fields such as river networks~\cite{tarboton1988fractal}, mobile networks, population genetics~\cite{berestycki2009recent}, and explosive percolation~\cite{achlioptas2009explosive,d2019explosive}.

The most common framework for studying CCA is the deterministic mean-field Smoluchowski equation, which describes the time evolution of the average cluster-size distribution under aggregation (see Refs.~\cite{leyvraz2003scaling, aldous1999deterministic,krapivsky2010kinetic,WATTIS20061,handbook} for reviews). Information about the physical system, such as cluster shape~\cite{puthalath2023lattice} as well as the transport mechanism responsible for bringing clusters together in space, is encoded in the collision kernel, which specifies the rate of collisions between clusters of different masses. The Smoluchowski equation is solvable for the average mass distribution for three collision kernels: the constant kernel, where the collision rate is independent of cluster masses; the sum kernel, where it is the sum of the colliding masses; and the product kernel, where the rate is proportional to the product of the colliding masses. In the case of the product kernel, a sol--gel transition occurs at a finite scaled time. Beyond this transition time, mass is no longer conserved and the Smoluchowski equation breaks down. More results on the typical evolution of the product kernel are reviewed in Sec.~\ref{model}.

The Smoluchowski equation ignores stochastic effects and therefore provides no information about fluctuations or deviations from the typical evolution. In recent work, we developed a general formalism based on a path-integral approach~\cite{rajesh2024exact}, together with a complementary biased Monte Carlo algorithm~\cite{dandekar2023monte}, to study large deviations in cluster--cluster aggregation. Within this framework, the probability $P(M,N,t)$ that $N$ particles remain at time $t$ when starting from $M$ monomers at $t=0$ can be expressed as the minimization of an action functional. For a general collision kernel, this leads to
\be
\lim_{M \to \infty}\frac{-\ln P(M, M \phi, \tau/M)}{M} = f(\phi, \tau),
\label{ldf_def}
\ee
where $f(\phi,\tau)$ defines the large-deviation function (LDF). We note that this is consistent with the Smoluchowski equation which has a sensible solution only in the scaled variable $\tau=Mt$. Likewise, we have to work in the limit of $\phi=N/M$ fixed. Closed-form expressions for $f(\phi,\tau)$ were obtained for the constant and sum kernels~\cite{rajesh2024exact}. For the product kernel, the LDF was expressed as a minimization problem. Although numerical analysis revealed a singularity (a jump in the second derivative) in the product kernel LDF, a complete analytic characterization of the LDF remained out of reach. We note that the path-integral approach is fully general, and has since been extended to $k$-nary aggregation processes ($kA\to \ell A$)~\cite{rajesh2025exact}.

An alternative approach to understanding the LDF for the product kernel exploits its connection to the Erd\H{o}s--R\'enyi random graph. Product-kernel aggregation provides a dynamic representation of this graph model: starting from monodispersed initial conditions, the mass distribution at any fixed time $t$ coincides with the distribution of connected-component sizes in an Erd\H{o}s--R\'enyi graph with edge probability $p = t/N$, where $N$ is the number of vertices. In this mapping, coagulation events map to the appearance of edges between previously disconnected components, while the gelation transition coincides with the emergence of a giant component. This mapping allows large-deviation properties of cluster sizes to be studied equivalently within the static graph ensemble or the dynamic aggregation framework~\cite{andreis2021large,andreis2023large}.

Engel \textit{et al.}~\cite{engel2004large} developed a large-deviation theory for the number of connected components in Erd\H{o}s--R\'enyi random graphs. Their approach exploited a formal correspondence between exponential moments of the graph ensemble, when restricted to graphs with number of links proportional to number of vertices,  and the free energy of the mean-field $q$-state Potts model, yielding explicit expressions for the convex envelope of the large-deviation function (CELDF) for integer values of $q$. The extension of these results to real positive $q$ was assumed, but not supported by independent analytical or numerical evidence. Moreover, while this approach revealed the presence of singularities in the CELDF, it did not yield a complete characterization of the associated phase behavior, such as analytic phase boundaries or a full phase diagram. Finally, the reliance on a non-bijective mapping to the Potts model limits the applicability of this framework to more general questions in product-kernel aggregation.

In this paper, we start from the master equation for product-kernel aggregation and obtain an exact integral representation of $P(M,N,t)$, valid for arbitrary finite values of $M$, $N$, and $t$. From this representation, we derive exact expressions for the exponential moments $\langle p^N \rangle$ for integer $p$, whose asymptotic analysis yields the corresponding large-deviation behavior. Employing a replica conjecture—numerically validated through finite-$M$ scaling—we extend these results to real $p \geq 0$. This allows us to compute the exact CELDF for the product kernel for all $\phi = N/M$ and scaled times $\tau = tM$. Analysis of the singular structure of the CELDF reveals a rich phase diagram, consisting of regions where the underlying LDF is non-convex and regions where the transition is continuous, separated by a tricritical point. The phase boundaries separating these regimes are obtained analytically. In addition, we derive the asymptotic form of the LDF in the limit of small $\phi$.

The remainder of the paper is organized as follows. Section~\ref{model} introduces the model. In Sec.~\ref{intrep}, we present a detailed derivation of the integral representation of \( P(M,N,t) \), the probability of observing exactly \( N \) particles at time \( t \) given \( M \) monomers at \( t=0 \). Section~\ref{expmoments} provides a rigorous derivation of the exponential moment \( \langle p^N \rangle \) for integer values of \( p \). Numerical evidence supporting the replica calculation is presented in Sec.~\ref{numerical}. In Sec.~\ref{meannumber} we confirm that the exponential moment correctly reproduces the known result for the mean fraction of particles that survive at time $t$. In Sec.~\ref{singular}, we analyze the singularities of the exponential moment, both as a function of \( \tau \) for fixed \( p \) and as a function of \( p \) for fixed \( \tau \). Section~\ref{rate} discusses the derivation of the large deviation function from the exponential moment. The singular behavior of the LDF and the resulting phase diagram  are derived in Sec.~\ref{ldfsingular}. Section~\ref{perturbation} contains a perturbative expansion that describes well the behavior for small $\phi$. We end with a summary and discussion of our results in Sec.~\ref{summary}.

\section{Model and Known Results \label{model}}

Consider a collection of particles labeled by their masses, undergoing mass-conserving binary aggregation (also known as the Marcus--Lushnikov model~\cite{marcus1968stochastic,lushnikov2006gelation,lushnikov1973evolution,lushnikov1978coagulation}). The process can be expressed as 
\begin{equation}
	A_i+A_j \xrightarrow {\text{$K(i,j)$}} A_{i+j},
\end{equation}
where $A_i$ represents a cluster of mass $i$, and the collision kernel $K(i,j)$ is the rate at which two clusters of masses $i$ and $j$ aggregate. We consider a monodispersed initial condition of $M$ particles of mass one. In this paper, we consider the product kernel, where the rate of two particles aggregating is proportional to the product of their masses, $i.e$,

\begin{equation}
	K(m_1,m_2)=\lambda m_1 m_2.
\end{equation}
We set $\lambda=1$ to set the time scale. We will focus on the probability that exactly $N$ particles remain at time $t$, denoted by $P(M,N,t)$,  where $N=1,2,...M$.

For the product kernel, we now summarize known results for the typical evolution. Earlier works have focused mainly on $\langle N \rangle$, the average number of particles, 
\begin{equation}
    \langle N \rangle=\sum_{N=1}^M N P(M,N,t),
\end{equation}
and  $\langle N_m \rangle$, the average number of particles of mass $m$. 

A standard way of studying the average mass distribution is through the mean-field Smoluchowski equation which ignores fluctuations:
	\begin{align}
		\frac{d \langle n_m\rangle}{d\tau}&=\frac{1}{2}\sum_{m_1=1}^{m-1}m_1(m-m_1)\langle n_{m_1}\rangle \langle n_{m-m_1}\rangle \notag \\
		&\quad-\langle n_m\rangle\sum_{m_1=1}^{\infty}m m_1\langle n_{m_1}\rangle, 
		\label{smol}
	\end{align}
where
\begin{eqnarray}
	\tau&=&Mt,\\
	\langle n_m\rangle&=&\frac{\langle N_m \rangle}{M},
\end{eqnarray}
 and mass conservation implies $\sum_m m \langle n_m \rangle =1$. 
 
Equation.~\eqref{smol} can be solved by the generating function method. The generating function of $n_m$ satisfies the inviscid Burgers equation and is known to be~\cite{van1985dynamic}
\begin{equation}
    \langle n_m\rangle=\frac{m^{m-2}\tau^{m-1}e^{-m\tau}}{m!} \quad \text{$m=1,2,...$}.
    \label{smolsol}
\end{equation}
However, the above solution does not conserve mass for $\tau>1$ ($i.e$, $\sum_m m \langle n_m\rangle<1$ for $\tau>1$). But it was shown, based on an exact solution for typical mass distribution in a finite system [see Eq.~\eqref{lushnikov}] that the expression for $\langle n_m \rangle$ in Eq.~\eqref{smolsol} continues to also hold for finite masses for all times $\tau \geq 1$. Consequently, Eq.~\eqref{smolsol} is valid for all $\tau$. As $\tau$ approaches $1$, $\langle n_m \rangle$ changes from an exponential distribution to a power law distribution
\begin{equation}
	\langle n_m \rangle \approx \frac{m^{-5/2}}{\sqrt{2\pi}} \quad \text{$m\gg1$}, ~~\tau=1,
\end{equation}
and first and second moments, derived from Eq.~\eqref{smolsol} are ~\cite{krapivsky2010kinetic}
\begin{align}
	\frac{\langle N \rangle}{M}=1-\frac{\tau}{2}, \quad \text{$\tau\leq1$},
	\label{eq:avenumber}
\end{align}
\begin{equation}
	\langle m^2 \rangle=\frac{1}{1-\tau}, \quad \text{$\tau<1$}.
\end{equation}
The loss of mass as well as divergence of $\langle m^2 \rangle$ at $\tau=1$ is attributed to the appearance of a gel, a particle whose mass is a finite fraction of the total mass.

Significant progress beyond the Smoluchowski equation was made by Lushnikov~\cite{lushnikov2006gelation,lushnikov1973evolution,lushnikov1978coagulation}, who derived the exact mean mass distribution for a system with finite $M$ for all $\tau$, thus going beyond the gelling time $\tau=1$. The dynamics of the coagulating system is described through a master equation for the joint probability of occupation numbers of clusters of all masses, which is then encoded in a generating functional. For the multiplicative kernel, this generating functional satisfies a closed evolution equation that can be solved exactly by exploiting its algebraic structure and an analogy with second-quantized bosonic fields. The average mass distribution is obtained by extracting the coefficients from the generating functional, leading to an explicit expression given by
\begin{equation}
	\begin{split}
		\langle n_m \rangle &=
		\binom{M}{m} e^{\left(m^2 - 2mM + m\right)\tau/2M} \\
		&\quad \times \left(e^{\tau/M} - 1\right)^{m-1}
		F_{m-1}\!\left(e^{\tau/M}\right),
	\end{split}
\label{lushnikov}
\end{equation}
where $F_m(x)$ are Mallows-Riordan polynomials, which obey the  recursion relation
\begin{equation}
	F_m(x)=\sum_{k=1}^{m}\binom{m-1}{k-1}\sum_{i=0}^{k-1}x^i F_{k-1}(x)F_{m-k}(x),
	\label{recursion}
\end{equation} 
with $F_0(x)=F_1(x)=1$.
For finite $m$, corresponding to taking the limit $M \to \infty$ keeping $m$ finite,  the mass distribution in Eq.~\eqref{lushnikov} reduces to Eq.~\eqref{smolsol} for all $\tau$. Thus, Eq.~\eqref{smolsol} describes the mass distribution for $m\sim O(1)$ for all time ~\cite{lushnikov2005exact}.

The number density of particles, obtained by summing over $m$ in Eq.~\eqref{smolsol}, is then given in terms of the Lambert $W$ function~\cite{lushnikov2005exact} 
\begin{equation}
	\frac{\langle N \rangle}{M}=-\frac{1}{\tau}W_0\left(-\tau e^{-\tau} \right)
	-\frac{\tau}{2}\left[\frac{1}{\tau}W_0\left(-\tau e^{-\tau} \right)\right]^2.
\end{equation}
For $\tau<1$, $-\frac{1}{\tau}W_0\left(-\tau e^{-\tau} \right)=1$ and hence $\langle N \rangle/M=1-\tau/2 $. The result for $\tau<1$ coincides with Eq.~(\ref{eq:avenumber}), the solution of the Smoluchowski equation. Using Eq.~\eqref{lushnikov} for $m \sim M^0$ [or equivalently Eq.~\eqref{smolsol}] the mass of the gel particle $g(\tau)$, given by $1-\sum_m m \langle n_m \rangle$,  simplifies to
\begin{equation}
	g(\tau)=1+\frac{1}{\tau}W_0(-\tau e^{-\tau}).
\end{equation}
For $\tau<1$, $\frac{1}{\tau}W_0(-\tau e^{-\tau})=-1$, implying that $g(\tau)=0$.

We emphasize that both the Lushnikov solution and Smoluchowski equation only describe the average or typical mass distribution. Also, the gelation transition can be viewed as a  loss of dynamical stability of the Smoluchowski equation~\cite{ball2012collective,dedola2026gelation}.

As described in Sec.~\ref{sec:intro}, product kernel aggregation is related to the Erd\H{o}s--R\'enyi random graphs~\cite{andreis2021large,andreis2023large}. The CELDF for the random graph has been studied previously via a mapping to the Potts model~\cite{engel2004large}.  In Table~\ref{compare_parameters}, we provide the mapping between the three models and the relation between the different parameters. 
\begin{table}
	\caption{\label{tab:ex}Comparison of parameters of the random graph model, mean-field $q$-state Potts model and product kernel aggregation.}
	\label{compare_parameters}
	\begin{ruledtabular}
		\begin{tabular}{lll}
			Random graph & Potts & Aggregation \\
			\hline
			$\gamma$ & $\beta$ (inverse temperature) & $\tau$\\
			$s_0$ (largest cluster) & $s_0$ (order parameter) & $\zeta$ \\
			$q$ & $q$ & $p$ \\
			$N$ (\# of vertices) & $N$ (\# of spins) & $M$
		\end{tabular}
	\end{ruledtabular}
\end{table}

\section{Integral representation of $P(M,N,t)$ \label{intrep}}

In this section, we derive an exact integral representation of $P(M,N,t)$, the probability of having exactly $N$ particles remaining at time $t$. Let $\vec{N}(t)$ be the $M$-dimensional vector 
\begin{equation}
	\vec{N}(t)=\{N_1(t),N_2(t)...N_M(t)\},
\end{equation}
 where $N_i(t)$ is the number of clusters with mass $i$ at time $t$. The probability of a system being in configuration $\vec{N}$, denoted by  $P(\vec{N},t)$, evolves according to the master equation:
 \begin{equation}
 	\begin{split}
 		&\frac{dP(\vec{N},t)}{dt}=\frac{1}{2}\sum_{i,j}i j \Big[(N_i+1+\delta_{i,j})(N_j+1) \\
 		&\times P(\vec{N}+I_i+I_j-I_{i+j})-N_i(N_j-\delta_{i,j})P(\vec{N})\Big],
 	\end{split}
 	\label{master}
 \end{equation}
where $I_k$ is the $M$-dimensional vector whose $k$-th component is unity and remaining components  are zero. The first term in the right-hand side of Eq.~\eqref{master} enumerates all possible collisions that lead to the configuration $\vec{N}$, while the second term enumerates the collisions that take the system out of $\vec{N}$. $P(M,N,t)$ is related to $P(\vec{N},t)$ as
\begin{equation}
	P(M,N,t)=\sum_{\vec{N}}P(\vec{N},t)\delta \left(\sum_{i=1}^{M}N_i(t)-N\right).
	\label{prob}
\end{equation} 

By introducing annihilation and creation operators, it is possible to rewrite $P(M,N,t)$ in terms of a ``Hamiltonian'' $H$~\cite{rajesh2024exact}. The derivation of $H$ is straightforward, and for completeness, we reproduce it in Appendix~\ref{appen}. We obtain
\begin{equation}
	P(M,N,t)=\frac{1}{N!} \langle 0 \vert \left(\sum_{k=1}^{M}a_k\right)^N e^{-Ht}a_1^{\dagger M} \vert 0 \rangle,
	\label{probability}
\end{equation}
where the Hamiltonian is given by
\begin{equation}
	H=-\frac{1}{2}\sum_{k,l=1}^{\infty}kl \left(a_{k+l}^\dagger-a_k^\dagger a_l^\dagger\right)a_k a_l,
\end{equation}
and $a_k$ and $a_k^\dagger$ are the creation and annihilation operators for mass $k$ obeying the  canonical commutation relation, $[a_m,a_n^\dagger]=\delta_{m,n}$.  $H$ can be rewritten in terms of  the mass operator $\hat{M}$~\cite{rajesh2024exact} as
\begin{equation}
	H=H_0+\frac{1}{2}\hat{M}^2,
\end{equation}
where 
\begin{eqnarray}
		\hat{M}&=& \sum_{m=0}^{\infty} ma^\dagger_m a_m ,\\
		H_0&=&-\frac{1}{2}\sum_{k,l=1}^{\infty}kl a_{k+l}^\dagger a_k a_l-\frac{1}{2}\sum_{k=1}^{\infty}k^2 a_k^\dagger a_k .
\end{eqnarray}
 $H_0$ is linear with respect to the creation operators, a feature that we will exploit later: 
\begin{equation}
	H_0=\frac{1}{2}\sum_{m=1}a_m^\dagger \Phi_m(a),
\end{equation}
where 
\begin{equation}
	\Phi_m(a)=-\sum_{k,l=1}^{\infty}kl \delta_{m,k+l}a_k a_l-m^2a_m.
	\label{phi}
\end{equation}

An explicit calculation shows that $[H_0,\hat{M}]=0$. Therefore,
\begin{equation}
	e^{-Ht}=e^{-H_0t}e^{-\frac{1}{2}\hat{M}^2t}.
	\label{evol}
\end{equation}
Substituting Eq.~\eqref{evol} for $e^{-Ht}$ in Eq.~\eqref{probability} and using the fact that the state $a_1^{\dagger M} \vert 0 \rangle$ is an eigenstate of the mass operator $\hat{M}$ with the eigenvalue $M$ we obtain
\begin{equation}
	P(M,N,t)=\frac{e^{-\frac{1}{2}M^2t}}{N!} \langle 0 \vert \left(\sum_{k=1}^{\infty}a_k\right)^N e^{-H_0t}a_1^{\dagger M} \vert 0 \rangle.
    \label{probfinal}
\end{equation}

Note that $\langle 0 \vert e^{-\mu a_1^\dagger}=\langle 0 \vert$, and
\begin{equation}
	a_1^{\dagger M}=M! \times \text{coefficient of $\mu^M$ in $e^{\mu a_1^\dagger}$} .
\end{equation}
We can then rewrite $P(M,N,t)$ in Eq.~\eqref{probfinal} as
\begin{equation}
	\begin{split}
		&P(M,N,t)=\frac{M!}{N!}e^{-\frac{M^2t}{2}}\oint \frac{d\mu}{2\pi i \mu^{1+M}}\\ &\times \langle 0 \vert e^{-\mu a_1^\dagger}\left(\sum_{k=1}^{M}a_k\right)^N
		 e^{-H_0 t}e^{\mu a_1^\dagger} \vert 0 \rangle.
	\end{split}
\label{contour}
\end{equation}	
The replacement $\left(\sum_{k=1}^{\infty}a_k\right)^N\to \left(\sum_{k=1}^{M}a_k\right)^N$ inside the inner product of Eq.~\eqref{contour} is justified due to mass conservation.

To proceed further, use the fact that $e^{-\mu a_1^\dagger}$ acts on functions of $a$ as a shift operator, $i.e.$,
\begin{equation}
	e^{-\mu a_1^\dagger}f(a_1,a_2,...)=f(a_1+\mu,a_2,...)e^{-\mu a_1^\dagger}.
\end{equation}
By commuting $e^{-\mu a_1^\dagger}$ in Eq.~\eqref{contour} to the right we obtain,
 \begin{equation}
	\begin{split}
		&P(M,N,t)=\frac{M!}{N!}e^{-\frac{M^2t}{2}}\oint \frac{d\mu}{2\pi i \mu^{1+M}}\\ &\times \langle 0 \vert \left(\sum_{k=1}^{M}\left(a_k+\mu \delta_{k,1}\right)\right)^N  e^{-H_0^\mu t} \vert 0 \rangle,
	\end{split}
	\label{path}
\end{equation}
where
\begin{equation}
	H_0^\mu=\frac{1}{2}\sum_{m=1}^{\infty}a_m^\dagger \Phi_m(a+\mu \delta_1),
\end{equation}
where $\Phi_m(a+\mu \delta_1)$ is a shorthand notation for $\Phi_m(a_1+\mu, a_2, \ldots)$.

The evolution operator $e^{-H_0^\mu t}$ is split into a product of the evolution operators $e^{-H_0^\mu dt}$ for infinitesimal times $dt$,
 \begin{equation}
	\begin{split}
		\raisetag{40pt}
		&P(M,N,t)=\frac{M!}{N!}e^{-\frac{M^2t}{2}}\oint \frac{d\mu}{2\pi i \mu^{1+M}}\\ &\times \langle 0 \vert \left(\sum_{k=1}^{M}\left(a_k+\mu \delta_{k,1}\right)\right)^N  e^{-H_0^\mu dt}e^{-H_0^\mu dt}... e^{-H_0^\mu dt}\vert 0 \rangle.
	\end{split}
	\label{path1}
\end{equation}
 
Using the standard path integral construction, we insert identity operators $I$ for every infinitesimal evolution in terms of coherent states, $\vert z \rangle$ and their complex conjugates. The coherent state $\vert z \rangle$ and $I$ are defined as follows
\begin{align}
     \label{coh1}
	a_i \vert z \rangle &= z_i \vert z \rangle, \\
    \label{coh3}
	I &= \int \prod_i\frac{dz_i d \tilde{z_i}}{\pi}e^{-\sum_iz_i \tilde{z_i}}\vert z \rangle \langle z \vert.
\end{align}
Using Eqs.~\eqref{coh1} and ~\eqref{coh3} in Eq.~\eqref{path1}, $P(M,N,t)$ is written as 
\begin{equation}
	\begin{split}
		\raisetag{100pt}
		&P(M,N,t)= \frac{ e^{-\frac{M^2t}{2}}M!}{N!}\oint \frac{d\mu}{2\pi i \mu^{1+M}} \\
		&\times \int  \mathfrak{D}(z) \mathfrak{D}(z') \left[\mu+\sum_{k=1}^{M}z_k(t)\right]^N  \\
		&\times \exp \Biggl[-\int_{0}^{t}dt'
		\sum_{m=1}^{\infty}\tilde{z}_m\Bigl(
		\dot{z}_m 
+\frac{1}{2}\Phi_m(z+\mu \delta_1)
		\Bigr)
		\Biggr] \\
		&\times \exp\!\left(-\sum_{m}\tilde{z}_m(0)z_m(0)\right) ,
	\end{split}
\end{equation}

The integral over $\tilde{z}$ is straightforward, leading to delta functions:
\begin{equation}
	\begin{split}
		\raisetag{60pt}
		&P(M,N,t) = \frac{ e^{-\frac{M^2t}{2}}M!}{N!}
		\oint \frac{d\mu}{2\pi i\, \mu^{1+M}}
		\int \mathfrak{D}(z) \\
		&\times \left[\mu + \sum_{k=1}^{M} z_k(t)\right]^{N} 
        \prod_{ t'} \prod_{m=1}^{M}
		\delta\Bigl(
		\dot{z}_m 
		+ \tfrac{1}{2}\Phi_m\bigl(z+ \mu \delta_1\bigr)
		\Bigr) \\
		&\times \delta\bigl(z_m(0)\bigr).
	\end{split}
	\label{deltaintegral}
\end{equation}
Doing the integral over $z_m(t')$ leads to the differential equation satisfied by $z_m$
\begin{equation}
	\begin{split}
		\dot{z}_m+\frac{1}{2}\Phi_m(z+\mu\delta_1)&=0,\quad\forall m \\
		z_m(0)&=0, \quad \forall m.
	\end{split}
	\label{zmdiff}
\end{equation}

Let $\tilde{b}_k(t)$ be the solution of Eq.~\eqref{zmdiff}, then Eq.~\eqref{deltaintegral} becomes
\begin{equation}
	P(M,N,t)= \frac{ e^{-\frac{M^2t}{2}}M!}{N!}\oint \frac{d\mu}{2\pi i\mu^{1+M}} \left[\mu+\sum_{k=1}^{M}\tilde{b}_k(t)\right]^N.
\end{equation}
The functions $\tilde{b}_k(t)$ solve the following system of differential equations:
\begin{equation}
	\begin{split}
		&\dot{\tilde{b}}_k(t)=-\frac{1}{2}\Phi_k(\tilde{b}(t)+\mu \delta_1),\quad t>0,\quad k\geq 1,\\
		&\tilde{b}_k(0)=0,\quad k\geq 1.
	\end{split}
\end{equation}
Shifting the variable
\begin{equation}
	\tilde{b}_k+\mu \delta_{k,1} = b_k
\end{equation}
we arrive at
\begin{equation}
	P(M,N,t)=\frac{ e^{-\frac{M^2t}{2}}M!}{N!} \oint \frac{d\mu}{2\pi i\mu^{1+M}} \left(\sum_{k=1}^{M}b_k(t)\right)^N,
	\label{prob2}
\end{equation}
and
\begin{equation}
	\begin{split}
		&\dot{b}_k(t)=-\frac{1}{2}\Phi_k(\{b_k(t)\}),\quad t>0,\quad k\geq 1,\\
		&b_k(0)=\mu \delta_1,\quad k\geq 1,
	\end{split}
\end{equation}
where the function $\Phi_k$ is defined in Eq.~\eqref{phi}. To solve this set of differential equations, define the generating function for $b_k(t)$ to be
\begin{equation}
	G(x,t)=\sum_{k=1}^{\infty} x^k b_k(t).
	\label{genfcn}
\end{equation}
$G(x,t)$ solves the following initial value problem,
\begin{equation}
	\begin{split}
		&\dot{G}(x,t)=\frac{1}{2}\left(x\frac{\partial G}{\partial x}\right)^2+\frac{1}{2}x \frac{\partial}{\partial x}\left(x\frac{\partial G}{\partial x}\right), \\
		&G(x,0)=\mu x.
	\end{split}
\label{genfunction}
\end{equation}

Equation~\eqref{genfunction} is solved using standard Cole-Hopf transform: let $\psi_t(x)=\exp \left(G(x,t)\right)$. Then
\begin{equation}
	\begin{split}
		&\dot{\psi}_t(x)=\frac{1}{2} \left(x \partial_x\right)^2 \psi_t(x),\\
		&\psi_0(x)=\exp (\mu x).
	\end{split}
\label{psieqn}
\end{equation}
It can be checked that the solution to Eq.~\eqref{psieqn} is
\begin{equation}
	\psi_t(x)=\sum_{n=0}^{\infty} \frac{\left(x\mu\right)^n}{n!}e^{\frac{n^2t}{2}}.
	\label{eq:46}
\end{equation}
Equation~(\ref{eq:46}) exists as a formal power series, and has zero radius of convergence. Finally, we note that
\begin{equation}
	\sum_{k=1}^{M}b_k(t)=G^{(M)}(1,t)=\ln \psi_t^{(M)}(1),
\end{equation}
where the superscript $M$ denotes the projection of the formal power series onto the polynomials of degree $M$. Substituting this answer into Eq.~\eqref{prob2}, we obtain
\begin{equation}
	P(M,N,t)=\frac{ e^{-\frac{M^2t}{2}}M!}{N!} \oint \frac{d\mu}{2\pi i \mu^{M+1}}\left[\ln \psi_t^{(M)}(\mu)\right]^N,
	\label{prob27}
\end{equation}
where 
\begin{equation}
	\psi_t^{(M)}(\mu)=\sum_{n=0}^{M} \frac{\mu^n}{n!}e^{\frac{n^2 t}{2}}.
	\label{psi}
\end{equation}
Equation~\ref{prob27} is exact for all finite $M, N, t$.

We note that, $\ln \psi_t^{(M)}(\mu)$ can be expressed in terms of Mallows-Riordan polynomial (see Eqs.~\eqref{lushnikov},~\eqref{recursion}). The generating function for these polynomials is~\cite{alsmeyer2023persistence}
\begin{equation}
	\ln \left(\sum_{n=0}^{\infty}x^{\frac{n(n-1)}{2}}\frac{z^n}{n!}\right)=\sum_{n=1}^{\infty} (x-1)^{n-1}F_{n-1}(x)\frac{z^n}{n!},
\end{equation}
leading to 
\begin{equation}
	\ln \psi_t(\mu)=\sum_{n=1}^{\infty}\left(e^t -1\right)^{n-1}e^{\frac{nt}{2}}F_{n-1}\left(e^t\right)\frac{\mu^n}{n!}.
	\label{poly}
\end{equation}
From the definition of the generating function in Eq.~\eqref{genfcn}, we can then read out  $b_k$ to be
\begin{equation}
	b_k(t)=(e^t-1)^{k-1}F_{k-1}(e^t)e^{\frac{kt}{2}} \frac{\mu^k}{k!}.
\end{equation}

We now provide a few consistency checks for the correctness of $P(M, N, t)$.
\begin{enumerate}[(i)]
    \item \textit{Total probability should add up to 1.}
    \begin{equation}
	\begin{split}
		\sum_{N=0}^{\infty}P(M,N,t)&=M!e^{-\frac{M^2t}{2}}\oint \frac{d\mu}{2\pi i \mu^{1+M}}\psi_t^{(M)}(\mu)  \\
		&=M!e^{-\frac{M^2t}{2}} \frac{1}{M!}e^{\frac{M^2t}{2}} =1.
	\end{split}
    \end{equation}
    
 \item \textit{The number of particles cannot be greater than $M$, $i.e$, $P(M,N,t)=0$ for $N>M$ and there must be at least one particle, $i.e$, $P(M,0,t)=0$.}
    
 For small $\mu$ the leading term for $\ln \psi$ in  Eq.~\eqref{poly} is $\mathcal{O}(\mu)$. Hence, we can write
\begin{equation}
	\ln \psi_t^{(M)}=\mu S(\mu),
\end{equation}
where $S(\mu)$ is analytic in the neighborhood of zero. Hence,
\begin{align}
	P(M,N,t)&=\frac{ e^{-\frac{M^2t}{2}}M!}{N!} \oint \frac{d\mu}{2 \pi i \mu^{1+M-N}}\left(S(\mu)\right)^N \notag\\
	&=0, \quad \text{$N>M$},\\
	P(M,0,t)&=\frac{ e^{-\frac{M^2t}{2}}M!}{N!} \oint \frac{d\mu}{2 \pi i \mu^{1+M}} \notag \\
	&=0, \quad \text{$M>0$}.
\end{align}
   
   \item \textit{At time $t=0$, all particles are monomers, $i.e$, $P(M,N,0)=\delta_{M,N}$.}
\begin{equation}
	\begin{split}
		P(M,N,0)&=\frac{M!}{N!}\oint \frac{d\mu}{2\pi i \mu^{1+M}}\left(\ln \sum_{n=0}^{M}\frac{\mu^n}{n!}\right)^N  \\
		&=\frac{M!}{N!}\oint \frac{d\mu}{2 \pi i \mu^{1+M}} \left(\ln e^\mu\right)^N \\
		&=\frac{M!}{N!}\oint \frac{d\mu}{2 \pi i \mu^{1+M-N}} =\delta_{M,N}.
	\end{split}
\end{equation}
   
   \item \textit{When $N=M$, no aggregation events occur in the interval $\big[0,t\big),$ and hence $P(M,M,t)=e^{-\frac{M(M-1)t}{2}}$.}
   
   From Eq.~\eqref{poly},
\begin{equation}
	\ln \psi_t^{(M)}=\mu e^{t/2}H(\mu),
\end{equation}
where
\begin{equation}
	\begin{split}
		H(\mu)&=1+\frac{\mu}{2!}\left(e^t-1\right)e^{t/2}F_1(e^t)\\&+\frac{\mu^2}{3!}\left(e^t-1\right)^2 e^t F_2(e^t)+...
	\end{split}
\end{equation}
with $H(0)=1$. Hence 
\begin{equation}
	\begin{split}
		P(M,M,t)&=e^{-\frac{M^2t}{2}}\oint \frac{d\mu}{2 \pi i\mu}\left(H(0)e^{t/2}\right)^M, \\
		&=e^{-\frac{M(M-1)t}{2}}.
	\end{split}
\end{equation}
\end{enumerate}

\begin{figure}
	\includegraphics[width=\columnwidth]{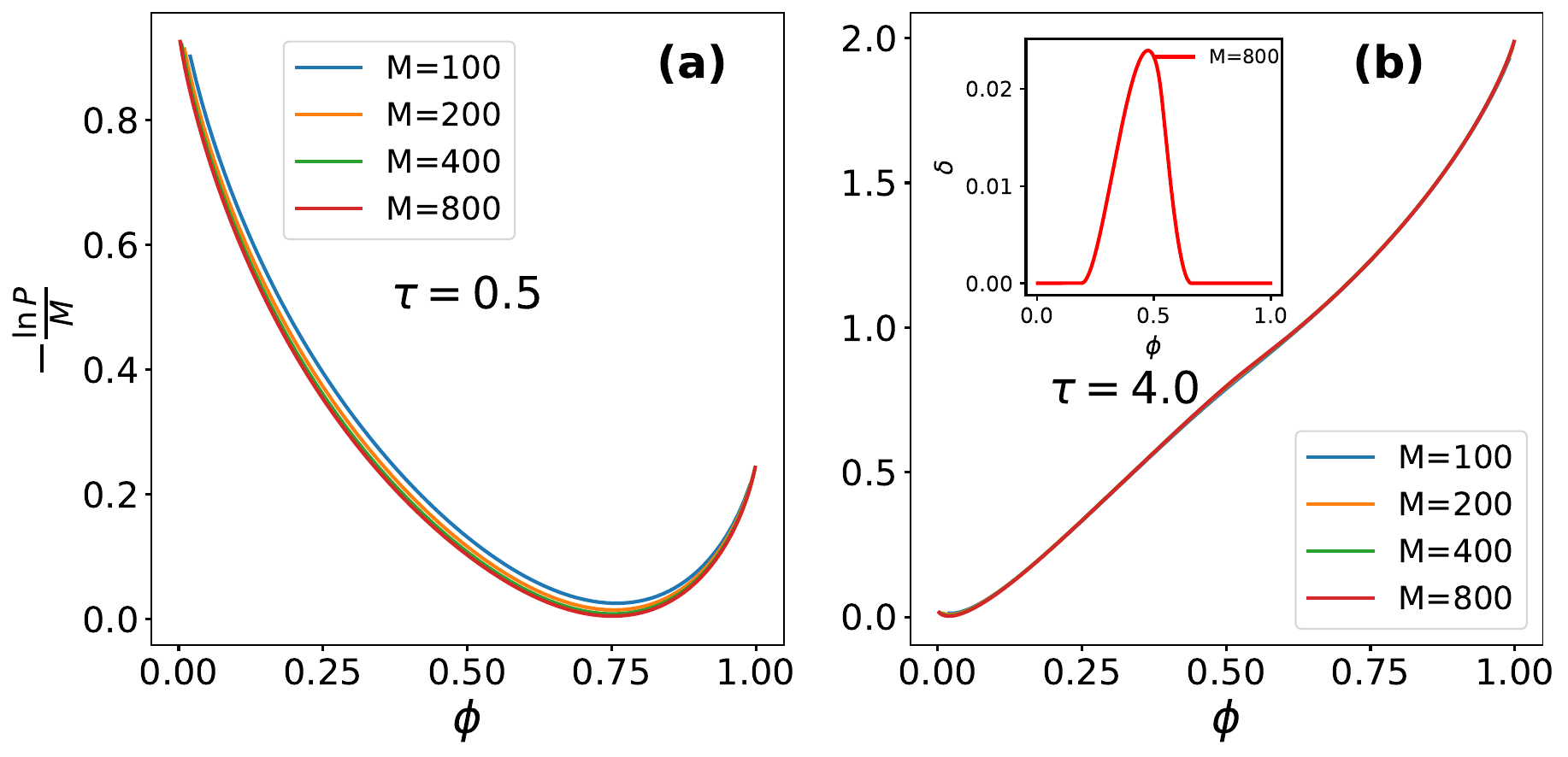}
	\caption{LDF obtained from the integral representation of $P(M,N,t)$ given in Eq.~\eqref{prob27}for $M$ using when (a) $\tau=0.5$ and (b) $\tau=4.0$. Inset: $\delta$, the difference between the LDF for $M=800$ and its CELDF.}
	\label{probability_integral}
\end{figure}

Using Eq.~\eqref{poly}, $\left(\ln \psi_t^{(M)}(\mu)\right)^N$ can be expanded as a power series in $\mu$ for any finite $M,N$ and the coefficient of $\mu^M$ can be extracted. Substituting into Eq.~(\ref{prob27}) allows us to determine $P(M, N, t)$ (the details of numerical evaluation is given in Appendix~\ref{numerical_app}).  The data for four different values of $M=100,200,400,800$ are shown in Fig.~\ref{probability_integral}(a) [$\tau=0.5$] and Fig.~\ref{probability_integral} (b) [$\tau=4.0$]. For $\tau=4.0$, a non-convexity is observed in the LDF, as can be clearly seen in the inset where the difference between the LDF and the CELDF for M=800 is shown. The detailed analysis of the LDF is presented in later sections. 

Though, Eqs.~\eqref{prob27} and~\eqref{poly} give an exact integral representation for $P(M,N,t)$ for any $M, N, t$, the presence of logarithm in Eq.~\eqref{prob27} together with a non-trivial $t$-dependent distribution of zeros of the polynomial $\psi_t^{(M)}(\mu)$ complicate the asymptotic analysis in the limit $M \rightarrow \infty$. Instead, we focus on the exponential moments.

\section{Exponential Moments \label{expmoments}}

To make analytical progress and to extract the large-deviation behavior, we now focus on the exponential moments $\langle p^N \rangle$ rather than the probability $P(M,N,t)$ directly. The advantage is that the moments can be evaluated exactly for integer $p$, and their asymptotic analysis can be carried out. Moreover, the convex envelope of the large-deviation function can be obtained from these moments via a Legendre-Fenchel transform. Starting from the exact integral representation \eqref{prob27}, we derive a closed expression for $\langle p^N \rangle$ that reduces, in the limit $M\to\infty$, to a $p$-fold Gaussian integral. Evaluating this integral by saddle point yields the scaled exponential moments $\lambda(\tau,p)$. The saddle-point equations undergo a symmetry-breaking transition: beyond a critical $\tau$, a non-symmetric solution emerges and dominates. This indicates a phase transition which we analyze in detail. Results for integer $p$ are extended to real $p \ge 0$ via a replica conjecture (Sec.~\ref{numerical}), and the convex large-deviation function is constructed in Sec.~\ref{rate}.

The exponential moment is obtained by multiplying Eq.~\eqref{prob27} by $p^N$ and summing over all $N$. Then,
\begin{equation}
	\langle p^N \rangle= e^{-\frac{M^2t}{2}}M! \oint \frac{d\mu}{2\pi i \mu^{M+1}}\left(\psi_t^{(M)}(\mu)\right)^p.
	\label{expmoment}
\end{equation}
Note that $\psi_t^{(M)}(\mu)$ is a sum over Gaussians [see Eq.~\eqref{psi}]. To transform them into exponentials, we introduce new variables using the standard integral
\begin{equation}
	e^{\frac{k^2t}{2}}=\int \frac{dx}{\sqrt{2 \pi t}}e^{-\frac{x^2}{2t}+kx}.
	\label{integral}
\end{equation}
Substituting Eq.~\eqref{integral} in each of the $p$ factors of $\psi_t^{(M)}$ in Eq.~\eqref{expmoment}, we obtain
 \begin{equation}
 	\begin{split}
 		\langle p^N \rangle &= e^{-\frac{M^2t}{2}}\sum_{k_1,k_2...k_p=0}^M \frac{M!}{k_1!k_2!...k_p!} \\
 		&\times \oint\frac{d\mu}{2\pi i\mu^{1+M}}\mu^{k_1+k_2+...k_p} \\
 		&\times \int \frac{dx_1dx_2...dx_p}{(2\pi t)^{p/2}} e^{\sum_{n=1}^{p}-(\frac{1}{2t}x_n^2-k_nx_n)}.
 	\end{split}
 	\label{expmoment2}
 \end{equation}
Carrying out the contour integral in Eq.~\eqref{expmoment2} leads to
\begin{equation}
	\begin{split}
		\raisetag{25pt}
		\langle p^N \rangle &= e^{-\frac{M^2t}{2}}\sum_{k_1,k_2...k_p=0}^M \frac{M!}{k_1!k_2!...k_p!} \delta \left(\sum_{i=1}^{p}k_i-M\right) \\
		&\times \int \frac{dx_1dx_2...dx_p}{(2\pi t)^{p/2}} e^{\sum_{n=1}^{p}-(\frac{1}{2t}x_n^2-k_nx_n)}.
		\label{expmoment3}
	\end{split}
\end{equation}
The summation over $k_i$  in Eq.~\eqref{expmoment3} is just the  multinomial expansion. Thus,
\begin{equation}
		\langle p^N \rangle = e^{-\frac{M\tau}{2}}\left(\sqrt{\frac{M}{2\pi \tau}}\right)^p \int \prod_{k=1}^{p} \left[dx_k e^{-\frac{M x_k^2}{2\tau}} \right] \left[\sum_{k=1}^{p}e^{x_k}\right]^M\!\!\!\!,
	\label{exp1}
\end{equation}
where $\tau=Mt$ is scaled time. 

We define the scaled exponential moment as
\begin{equation}
	\lambda(\tau,p)=\lim_{M \rightarrow \infty} \frac{1}{M} \ln \langle p^N \rangle.
	\label{expmoment_def}
\end{equation}
 Evaluating the integral \eqref{exp1} using the saddle-point approximation and taking the limit $M \rightarrow \infty$, we obtain
\begin{equation}
	\lambda(\tau,p) =\sup_{\{x\}}\{F(\tau,\{x\})-\frac{\tau}{2}\},
\end{equation}
where 
\begin{equation}
	F(\tau,\{x\})=-\frac{1}{2 \tau}\sum_{k=1}^{p}x_k^2+\ln \sum_{k=1}^{p}e^{x_k}.
\end{equation}

Notice that $\{x\} \to \pm \infty$, $F(\tau,\{x\})=-\infty$. Therefore, the supremum of $F(\tau,\{x\})$ is achieved at a finite critical point. The critical point, obtained by setting the derivative with respect to $x_k$ to zero, is given by
\begin{equation}
	x_k=\frac{\tau}{Z}e^{x_k}, \quad \text{$k=1,2,...p$},
	\label{critic}
\end{equation}
where $Z=\sum_{k=1}^{p}e^{x_k}$. It follows from Eq.~\eqref{critic} that $\sum_{k=1}^{p}x_k=\tau$. Therefore, Eq.~\eqref{critic} can be equivalently expressed in terms of $p+1$ variables $x_1,x_2,...x_p,Z$ to be
\begin{align}
	\sum_{k=1}^{p}x_k&=\tau,\label{norm} \\
	x_ke^{-x_k}&=\frac{\tau}{Z}, \quad k=1,2,\dots,p	\label{crit}.
\end{align}

One of the solutions to Eqs.~\eqref{norm} and ~\eqref{crit} is
\begin{align}
	x_k^{(0)}&=\frac{\tau}{p}, \quad k=1,2,...,p, 	\label{x0} \\ Z&=pe^{\tau/p}.
\end{align}
 The corresponding value of the function $F(\tau,\{x\})$ is
\begin{equation}
	F_0^{(c)}(\tau,p)=\frac{\tau}{2p}+\ln p.
	\label{class}
\end{equation}

\begin{figure}
	\includegraphics[width=\columnwidth]{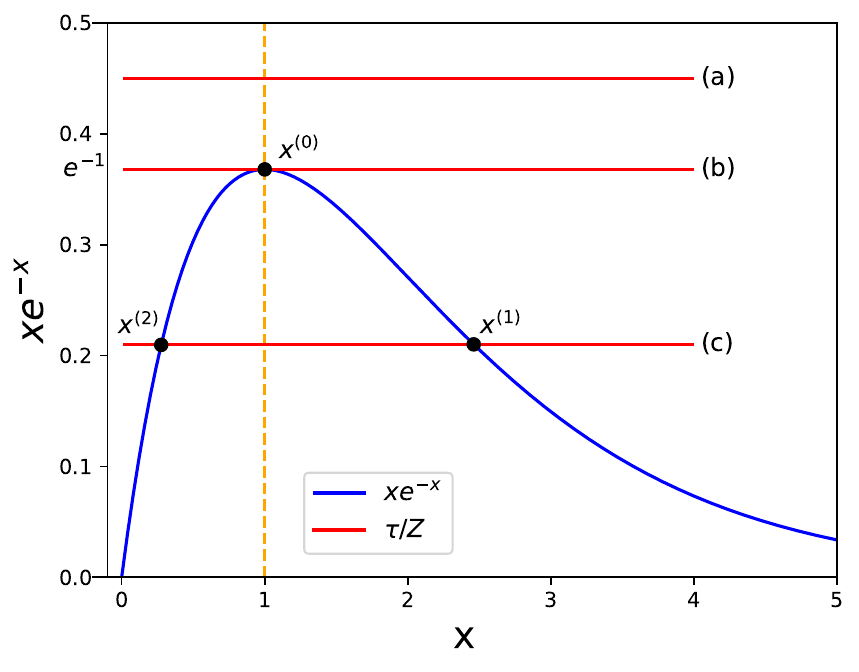}
	\caption{The possible solutions of Eq.~\eqref{crit} for $x$, depending on the values of $\tau/Z$. (a) No solutions when $\tau/Z>e^{-1}$, (b) one solution $x^{(0)}=1$ when $\tau/Z=e^{-1}$ and (c) two distinct solutions $x^{(1)}$ and $x^{(2)}$ when $\tau/Z<e^{-1}$.} 
	\label{solutions}
\end{figure}

We now look for other solutions. The function $xe^{-x}$ has a maximum of $e^{-1}$ at $x=1$ (see Fig.~\ref{solutions}). Hence, depending on the value of $\tau/Z$, Eq.~\eqref{crit} can have either two, one, or zero solutions, as illustrated in Fig.~\ref{solutions}.

Consider the case when one or two solutions for $x$ are possible ($\tau/Z\leq e^{-1}$). We denote them as $x^{(1)}$ and $x^{(2)}$ with $x^{(1)}\geq x^{(2)}$, as indicated in Fig.~\ref{solutions}. Since the maximum of $xe^{-x}$ occurs at $x=1$, we obtain
\begin{equation}
	x^{(2)} \leq 1 \leq x^{(1)}.
	\label{x1x2}
\end{equation}
Since $x^{(1)} \geq 1$ and $x^{(2)} \geq 0$, Eq.~\eqref{norm} implies that for $\tau<1$, the only possible solution is Eq.~\eqref{x0}, where all $x_k$ are equal.

We now consider $\tau\geq 1$. Let $m_1$ be the multiplicity of $x^{(1)}$ and $m_2$ be the multiplicity of $x^{(2)}$. Clearly, $m_1+m_2=p$. Using Eq.~\eqref{norm}, the solution for $x$ can be parameterized as follows:
\begin{equation}
	\begin{aligned}
		x^{(1)}&=\frac{\tau}{p}\left(1+m_2\zeta\right), \\
		x^{(2)}&=\frac{\tau}{p}\left(1-m_1\zeta\right).
	\end{aligned}
\label{para}
\end{equation}
Then $x^{(1)}-x^{(2)}=\tau \zeta$. Since $x^{(1)}\geq x^{(2)}$ (see Eq.~\eqref{x1x2}) we have the condition $\zeta \geq 0$. As $x^{(1)}=\tau e^{x^{(1)}} /Z$  and $x^{(2)}=\tau e^{x^{(2)}}/Z$ taking the ratio $x^{(1)}/x^{(2)}$ in Eq.~\eqref{para} we obtain the equation satisfied by $\zeta$ to be
\begin{equation}
	\zeta=\frac{e^{\tau \zeta}-1}{m_2+m_1e^{\tau \zeta}}.
	\label{zeta}
\end{equation}

$\zeta=0$ is always a solution of Eq.~\eqref{zeta}, corresponding to all $x_k$ being equal to $\tau/p$, with the associated value of $F(\{x\},\tau)$ given by Eq.~\eqref{class}.  Depending on the values of $m_1$, $m_2$, and $\tau$, $\zeta$ can have up to two non-zero solutions (see below after Eq.~\eqref{eigenvalues}).

We then have to find the critical point corresponding to the global maximum of $F_\tau(x)$. At the point of a local maximum $x_c$ the Hessian matrix $\nabla \otimes \nabla F_\tau(x_c)$ is negative definite. An explicit computation gives,
\begin{equation}
	\partial_i \partial_j F_\tau(x)=\frac{x_i-1}{\tau} \delta_{i,j}-\frac{x_ix_j}{\tau^2}, \hspace{0.3cm} 1 \leq i,j \leq p.
	\label{hessian}
\end{equation}
The eigenvalue $\Lambda$ of the Hessian satisfies the characteristic equation 
\begin{equation}
	\begin{split}
		&\left(\frac{x^{(1)}}{\tau}-\frac{1}{\tau}-\Lambda\right)^{m_1-1}
		\left(\frac{x^{(2)}}{\tau}-\frac{1}{\tau}-\Lambda\right)^{m_2-1}\\
		\times 
		&\bigg[
		\left(\frac{x^{(1)}}{\tau}-\frac{1}{\tau}-\Lambda\right)
		\left(\frac{x^{(2)}}{\tau}-\frac{1}{\tau}-\Lambda\right)
		\\&-m_1\frac{x^{(1)^2}}{\tau^2}\left(\frac{x^{(2)}}{\tau}-\frac{1}{\tau}-\Lambda\right)
		\\&-m_2\frac{x^{(2)^2}}{\tau^2}\left(\frac{x^{(1)}}{\tau}-\frac{1}{\tau}-\Lambda\right)
		\bigg]=0,
	\end{split}
\label{characteristic}
\end{equation}
which is easily solved to give 
\begin{equation}
	\begin{aligned}
		\Lambda_1&=\frac{x^{(1)}-1}{\tau} \quad\text{(Multiplicity $m_1-1$),}\\
		\Lambda_2&=\frac{x^{(2)}-1}{\tau} \quad\text{(Multiplicity $m_2-1$),}\\
		\Lambda_3&=-\frac{1}{\tau} \quad\text{(Multiplicity 1),} \\
		\Lambda_4&=\frac{px^{(1)}x^{(2)}-\tau}{\tau^2} \quad \text{(Multiplicity 1)}.
	\end{aligned}
	\label{eigenvalues}
\end{equation}

Since $x^{(1)}\geq 1$ [see Eq.~\eqref{x1x2}], $\Lambda_1 \geq 0.$ Since all eigenvalues are negative at the critical point, we conclude that a local maximum must correspond to $m_1=1$. Since $x^(2) <1$ [see Eq.~\eqref{x1x2}], we obtain $\Lambda_2 \leq 0$ for all times whereas $\Lambda_3<0$ automatically. The condition that $\Lambda_4$ is negative is equivalent to
\begin{equation}
	\frac{\tau}{p}\left(1+m_2\zeta\right)\left(1-m_1\zeta\right) \leq 1,
	\label{inequality}
\end{equation}
where we can assume $m_1=1$. We note that the critical point in Eq.~\eqref{x0} is a special case of the more general critical point in Eq.~\eqref{para}, obtained when $\zeta = 0$.

We now examine the non-zero solutions to Eq.~\eqref{zeta}, which corresponds to the zeros of the function
\begin{equation}
	g(\zeta)=\zeta - \frac{e^{\tau \zeta} - 1}{ e^{\tau \zeta}+p-1},
	\label{zeta-final}
\end{equation}
which is obtained from Eq.~\eqref{zeta} with $m_1=1$ and $m_2=p-1$. Since the second term in the right hand side of Eq.~\eqref{zeta-final} is less than $1$ for any $p>0$, it is clear that $g(\zeta)>0$, for all $\zeta \geq 1$. Thus, all roots of $g$ lie in $[0,1)$. We also note that $g'(\zeta)\vert_{\zeta=0}=1-\tau/p$ changes sign at $\tau=p$.

\begin{figure}
	\includegraphics[width=\columnwidth]{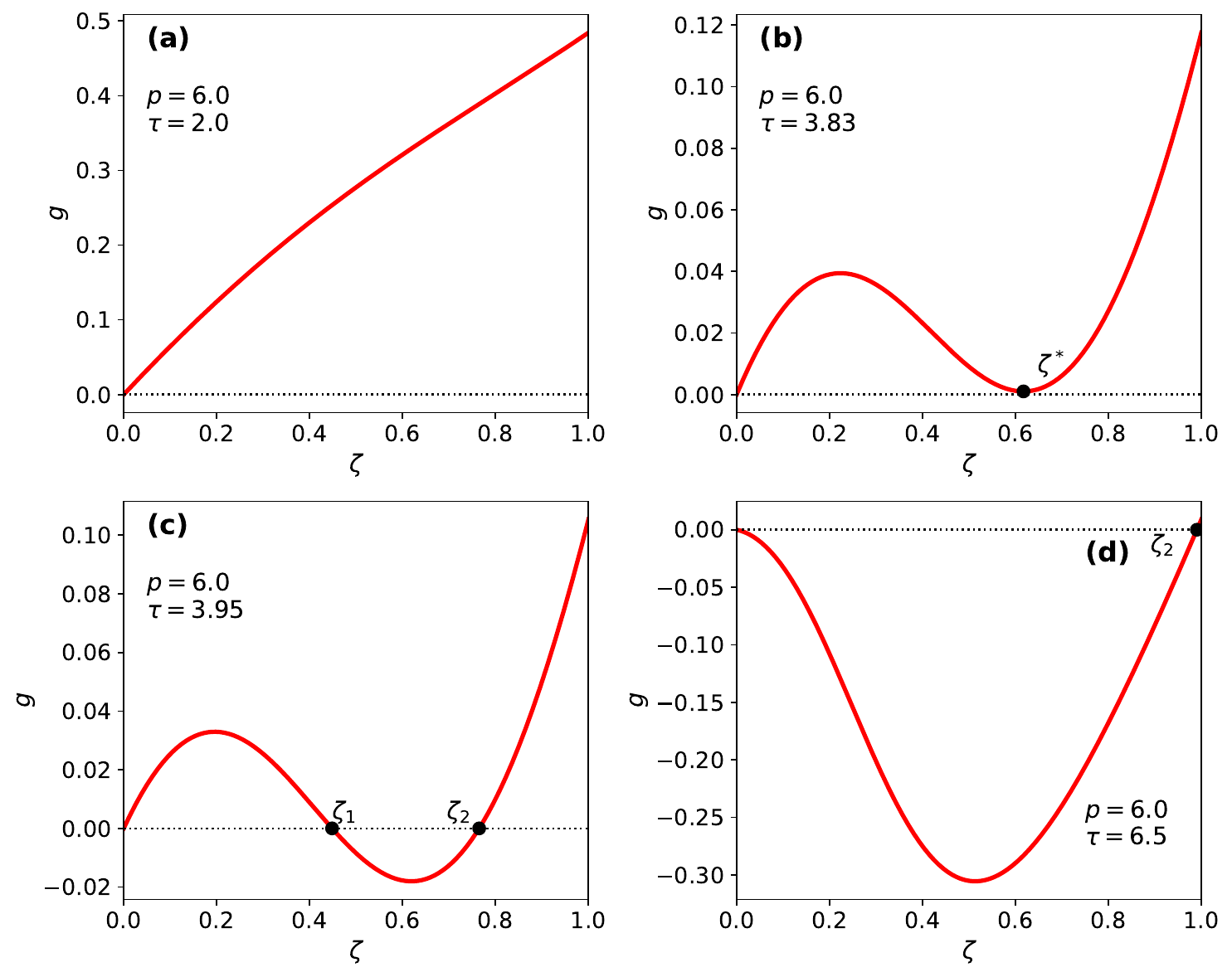}
	\caption{The behavior of $g(\zeta)$ for different values of $\tau$ for fixed $p=6.0$ for (a) $\tau = 2 $; $\zeta = 0$ is the only root; (b) $\tau = 3.83\dots = \tau^*$; when there is a double root at $\zeta^* = 0.62$; (c) $\tau = 3.95>\tau^*$; when there are $\zeta_1$ and $\zeta_2$; and (d) $\tau = 6.5 > p$; when there is only one non-zero root.}
	\label{zeta_solutions}
\end{figure}

The variation of $g(\zeta)$ with $\zeta$ for different $\tau$ and fixed $p=6.0$ is shown in Fig.~\ref{zeta_solutions}. For small $\tau$, there is only one root $\zeta=0$ [see Fig.~\ref{zeta_solutions}(a)]. As $\tau$ is increased, a double root appears, as shown in Fig.~\ref{zeta_solutions}(b) (we will denote the value of $\tau$ as $\tau^*$). Beyond $\tau^*$, there are $3$ roots [see Fig.~\ref{zeta_solutions}(c)]. Once $\tau$ crosses $p$, the slope at $0$ changes sign and there are only two roots [see Fig.~\ref{zeta_solutions}(d)]. The phenomenology is summarized as

\begin{enumerate}[(i)]
	\item For $\tau < \tau^*$, $\zeta = 0$ is the only root.
	\item When $\tau = \tau^*$, the roots are $\zeta = 0$ and a double root $\zeta^*$.
	\item For $\tau^* < \tau < p$, there are three roots: $\zeta = 0$, $\zeta_1$, and $\zeta_2$.
	\item For $\tau > p$, there are two roots $\zeta = 0$ and $\zeta_2$.
\end{enumerate}

At $\zeta^*$, where a double root occurs, both $g(\zeta^*) = 0$ and $g'(\zeta^*) = 0$ [see Fig.~\ref{zeta_solutions}(b)]. This leads to
\begin{align}
	&\zeta^* - \frac{e^{\tau^* \zeta^*} - 1}{p - 1 + e^{\tau^* \zeta^*}} = 0,\label{zetastar}\\
	&1+\frac{e^{\zeta^*  \tau^* } \left(-1+e^{\zeta^*  \tau^* }\right) \tau^* }{\left(e^{\zeta^*  \tau^* }+p-1\right)^2}-\frac{e^{\zeta^*  \tau^* } \tau^* }{e^{\zeta^*
			\tau^* }+p-1}=0 \label{zetastar_derivative}.
\end{align}
Eliminating $e^{\tau^* \zeta^*}$ from Eqs.~\eqref{zetastar} and ~\eqref{zetastar_derivative}, we obtain
\begin{equation}
	\frac{\tau^*}{p}\left(1+(p-1)\zeta^*\right)\left(1-\zeta^*\right) = 1,
\end{equation}
which falls at the edge of the bound in Eq.~\eqref{inequality}.

For $\tau^* < \tau < p$, there exist three roots: $0$, $\zeta_1$, and $\zeta_2$. For $\zeta_1$, $\zeta_2$ Eq.~\eqref{zetastar} still holds, while $g'(\zeta_1)<0$ and $g'(\zeta_2)>0$ [see fig.~\ref{zeta_solutions}(c)]. Consequently, Eq.~\eqref{zetastar_derivative} is modified to
\begin{equation}
	1+\frac{e^{\zeta_1  \tau } \left(-1+e^{\zeta_1  \tau }\right) \tau }{\left(-1+e^{\zeta_1  \tau }+p\right)^2}-\frac{e^{\zeta_1  \tau } \tau }{p-1+e^{\zeta_1
			\tau }}<0,
	\label{zeta1-derivative}
\end{equation}
with $<$ replaced by $>$ for $\zeta_2$. Substituting for $e^{\tau \zeta_1}$ from Eq.~\eqref{zetastar} leads to the inequality
\begin{equation}
	\frac{\tau}{p}\left(1+(p-1)\zeta_1\right)\left(1-\zeta_1\right) > 1,
\end{equation} 
which violates Eq.~\eqref{inequality}. Hence, $\zeta_1$ is a local minimum and cannot maximize $F$. By the same argument $\zeta_2$ is always a local maximum.

$F$ is maximized by either $\zeta=0$ or $\zeta=\zeta_2$. The exponential moment is given by
\begin{equation}
\lambda(\tau,p,\zeta)=\max_{\zeta \geq 0} F^{(c)}(p,\tau,\zeta)-\frac{\tau}{2}.
\label{replica}
\end{equation}
Here 
\begin{equation}
	\begin{split}
		F^{(c)}(\tau,p,\zeta)&=\frac{\tau}{2p}\left[1+(p-1)\zeta\right]\left[1-\zeta\right]+\ln p  \\
		&\quad-\frac{1}{2}\ln \left[\left(1+(p-1)\zeta\right)\left(1-\zeta\right)\right],
	\end{split}
	\label{F_general}
\end{equation}
and the maximum in Eq.~\eqref{replica} is taken over all non-negative solutions to Eq.~\eqref{zeta} for $m_1=1$. The exponential moment corresponding to $\zeta=0$ is
\begin{equation}
	\lambda(\tau,p,\zeta=0)=\frac{\tau}{2p}+\ln p-\frac{\tau}{2}.
	\label{classical}
\end{equation}
 We refer to this as the classical solution.
 
There exists a transition time, denoted by $\tau_c$, up to which the classical solution remains valid, and beyond which $\zeta_2$ become dominant.   Since the exponential moment is continuous~\cite{touchette2009large}, we obtain the condition at the transition point:
 \begin{equation}
 	F^{(c)}(\tau,p,\zeta)=F^{(c)}_0(\tau,p),
 	\label{trancondition}
 \end{equation}
which reduces to
\begin{equation}
	\left(1+(p-1)\zeta\right)\left(1-\zeta\right)=1.
	\label{zetarelation}
\end{equation}
Equation~\eqref{zetarelation} has two solutions: $\zeta_c=0$ and 
\begin{equation}
	\zeta_c=\frac{p-2}{p-1}.
	\label{zetac}
\end{equation}
When $\zeta_c = 0$, Eq.~\eqref{trancondition} is satisfied if all $x_k$ equal unity, which leads to $\tau_c = p$. For $p \leq 2$ the only possible solution is $\zeta_c=0$. Therefore we assume this relation to hold for $p \leq 2$. When $\zeta_c$ is as in Eq.~\eqref{zetac}, then substituting in to ~\eqref{zeta}, we obtain $\tau_c$ for $p >2$. Combining both results we obtain
\begin{equation}
	\tau_c =
	\begin{cases}
		\dfrac{2(p-1)}{p-2}\ln(p-1), & p> 2, \\
		p, & p \leq 2.
	\end{cases}
	\label{tauc}
\end{equation}

\section{Replica Conjecture \label{numerical}}

We now generalize the results obtained in Sec.~\ref{expmoments} for the exponential moments for integer $p\geq 2$ to real $p \geq 0$ -- an assumption we refer to as the replica conjecture. In particular, we generalize the exponential moment for any $p>0$ to be $F=F_0^{(c)}$, as in Eq.~\eqref{class}, for $\tau<\tau_c$ and $F=F^{(c)}(\tau,p,\zeta_2)$ as in Eq.~\eqref{F_general}, for $\tau\geq \tau_c$, where $\tau_c$ is given by Eq.~\eqref{tauc}. It turns out that for $0<p<1$, the analytically continued solution minimizes $F$. This is analogous to the replica trick in spin glasses and also the $O(n)$ model~\cite{griffiths1983convexity}.    

We now present numerical evidence in support of the results in Sec.~\ref{expmoments}, generalized to real $p$. To do so, we compare two results for exponential moments: the replica result for infinite $M$ given in Eq.~\eqref{replica}, and the exact expression for finite $M$ given in Eq.~\eqref{expmoment}. Consider Eq.~\eqref{expmoment}, in which $\langle p^N \rangle$ is expressed in terms of a polynomial $\psi_t^{(M)}$, for finite $M$. For a given value of $p$, $\left(\psi_t^{(M)}\right)^p$ can be expanded as a series in $\mu$ using the Mallows-Riordan polynomials and the coefficient of $\mu^M$ can be extracted (details of numerical evaluation are given in Appendix~\ref{numerical_app}).
  
For fractional values of $p$, use the fact that
\begin{equation}
	\left(\psi_t^{(M)}\right)^p=\exp \left(ph(\mu)\right),
\end{equation}
 where $g(\mu)=\ln \psi_t^{(M)}$, which is given in terms of Mallows-Riordan polynomials (see Eq.~\eqref{poly}). By Taylor expanding the exponential we get 
\begin{equation}
	\left(\psi_t^{(M)}(\mu)\right)^p=1+ph(\mu)+\frac{p^2 h^2(\mu)}{2!}+...\frac{p^Mh^M(\mu)}{M!}.
	\label{frac}
\end{equation}
Each term in the above expression can be expanded separately to obtain the exponential moment for any $M$.

The comparison of exponential moments obtained from the replica calculation given by Eq.~\eqref{replica} and the exact expression for finite $M$ given by Eq.~\eqref{expmoment} is done for three regions of $p$: $p>2$, $1 \leq p \leq 2$ and $p<1$. For comparison, we choose $p=2.5$, $p=1.5$, $p=0.5$ corresponding to one generic point for each of the above regions.

\begin{figure*}
	\includegraphics[width=0.8 \textwidth]{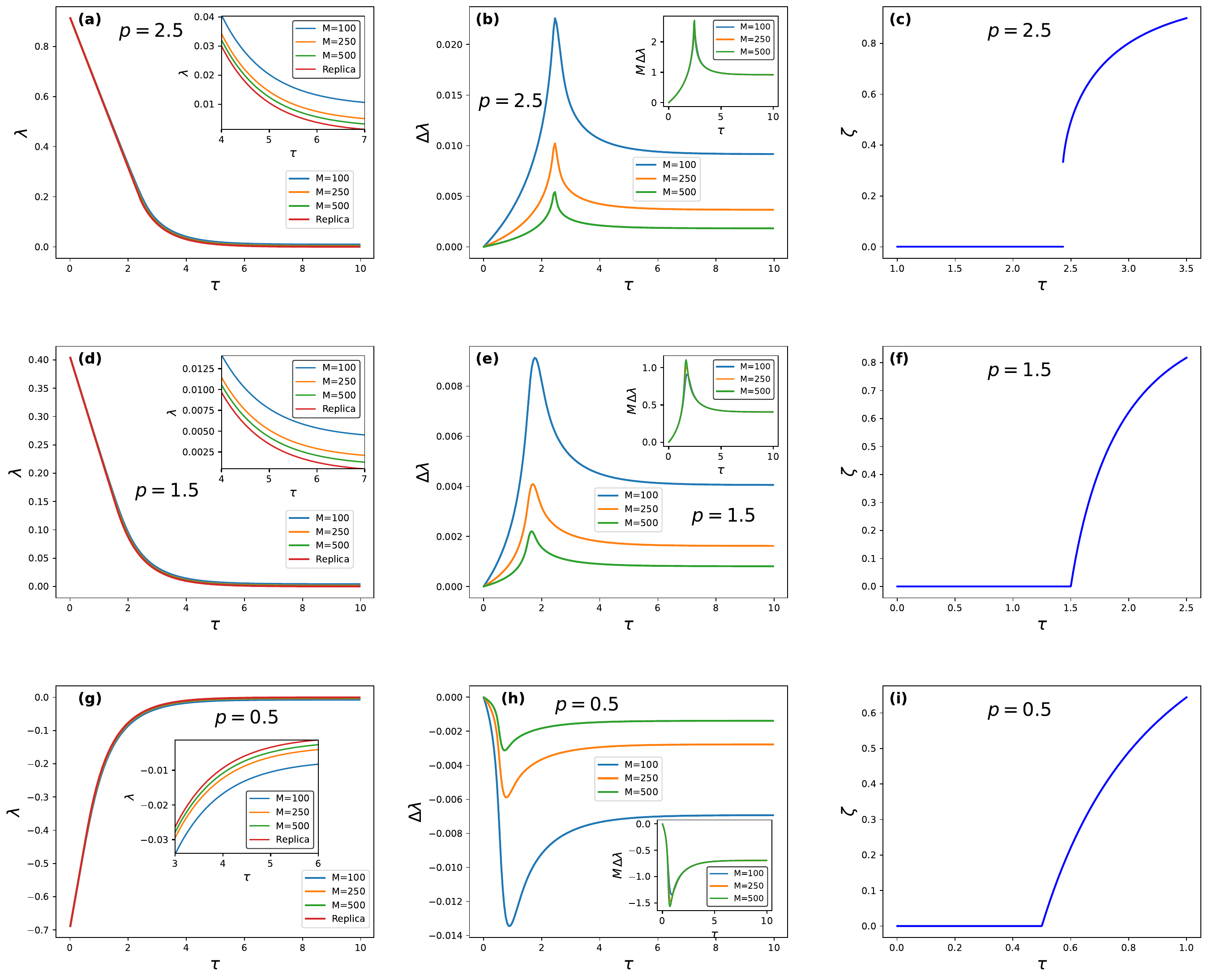}
	\caption{Comparison of the exponential moment obtained from replica calculation (Eq.~\eqref{replica}) for infinite $M$ with the series expansion (Eq.~\eqref{expmoment}) for finite $M$, when (a) $p=2.5$, (d) $p=1.5$ and (g) $p=0.5$. The difference between the exponential moment obtained from series expansion and replica calculation, denoted by $\Delta \lambda$ are shown in  (b) $p=2.5$, (e) $p=1.5$ and (h) $p=0.5$.  The corresponding insets show the data collapse  when plotted against $M \Delta \lambda$. The variation of $\zeta$ with $\tau$  for (c) $p=2.5$, (f) $p=1.5$ and (i) $p=0.5$. The transition time $\tau_c$ is identified as the smallest $\tau$ for which $\zeta$ is non-zero.}
	\label{comparison}
\end{figure*}

The data for three different values of $M=100,250,500$ are compared with the replica result in Fig.~\ref{comparison}(a) [$p=2.5$], Fig.~\ref{comparison}(d) [$p=1.5$] and Fig.~\ref{comparison}(g) [$p=0.5$]. While the data for different $M$ curves are close to the replica result, there are discrepancies as can be seen in the corresponding insets, as well as in Fig.~\ref{comparison}(b),(e),(h), where the differences with the replica result are shown. We show that these differences are only finite-size corrections, and vanish in the limit $M\to \infty$ by demonstrating that the difference $\Delta \lambda$ for different $M$ collapses onto a single curve when scaled with $M$, $i.e$,
\begin{equation}
	\Delta \lambda(\tau)=\frac{1}{M}g(\tau),
\end{equation}
where $g(\tau)$, a scaling function, is shown in inset of Fig.~\ref{comparison}(b),(e),(h). We then conclude that the replica result is correct for all real $p$.

\begin{figure}
	\includegraphics[width=0.8\columnwidth]{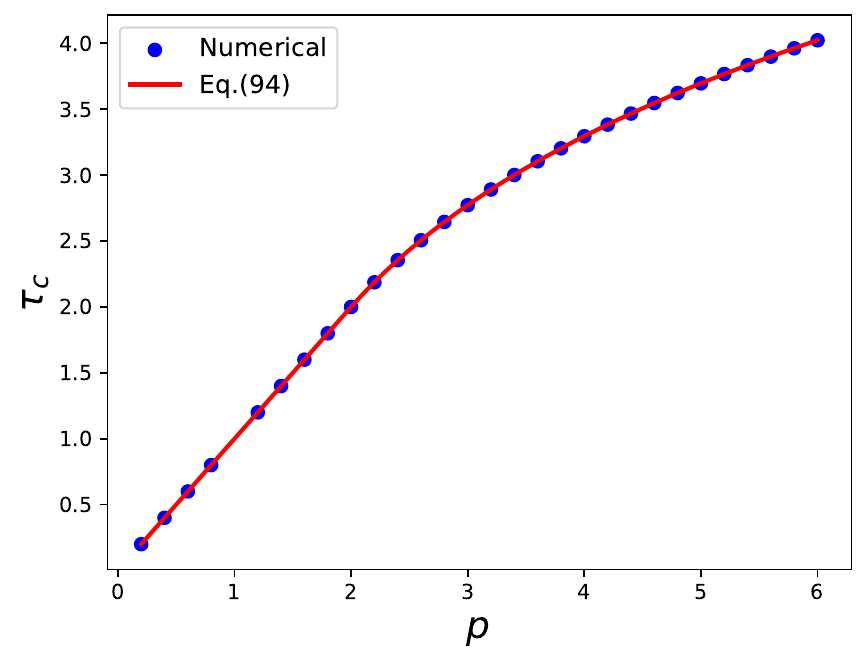}
	\caption{Comparison of the transition time $\tau_c$, as given in Eq.~\eqref{tauc}, with the $\tau_c$ obtained from the numerical solution of $\zeta$ with $\tau$ as shown in Fig.~\ref{comparison}(c), (f), (i). }
	\label{transition}
\end{figure}

We also check the correctness of the formula Eq.~\eqref{tauc} for the critical $\tau_c$. To do, we obtain $\tau_c$ numerically as follows. For fixed $p$, for each value of $\tau$, we determine the value of $\zeta$ that maximizes the expression in Eq.~(\ref{replica}). The variation of $\zeta$ with $\tau$ are shown in Fig.~\ref{comparison}(c) [$p=2.5$], Fig.~\ref{comparison}(f) [$p=1.5$] and Fig.~\ref{comparison}(i) [$p=0.5$]. The transition time $\tau_c$ is identified with the lowest value of $\tau$ where $\zeta$ deviates from zero. Clearly, for $p=2.5$, there is a jump in $\zeta$ while $\zeta$ is continuous for $p=1.5,0.5$. The numerically obtained $\tau_c$ is compared with Eq.~\eqref{tauc} in Fig.~\ref{transition}, and we see an excellent agreement.

\section{Mean Fraction of Particles}
\label{meannumber}

We now compute the mean number of particles $\langle N \rangle$ from the result for exponential moments.
\begin{equation}
    \frac{\langle N \rangle}{M}=p\frac{d\lambda(\tau,p)}{dp}\bigg |_{p=1},
\end{equation}
where
\begin{equation}
	\begin{split}
		\frac{d\lambda}{dp}&=\frac{dF_0^{(c)}}{dp} \quad \tau\leq 1 \\
		&=\frac{\partial F^{(c)}}{\partial p}+\frac{\partial F^{(c)}}{\partial \zeta}\frac{d \zeta}{d p} \quad \tau>1.
	\end{split}
\end{equation}
Therefore for $\tau \leq 1$, when $\zeta=0$, from Eq.~\eqref{class},
\begin{equation}
	\frac{\langle N \rangle}{M}=1-\frac{\tau}{2}, \quad \tau \leq 1.
\end{equation}

For $\tau>1$, from the critical equation~\eqref{zeta},
\begin{equation}
\frac{d\zeta}{dp}\bigg|_{p=1}=-\frac{(1-\zeta(1))\zeta(1)}{1-\tau(1-\zeta(1))}.
\end{equation}
and
\begin{equation}
	1-\zeta(1)=e^{-\tau\zeta(1)}.
\end{equation}
Using these results, it is easy to calculate the $p$-derivative of $F^{(c)}$. Introducing $\gamma=1-\zeta(1)$, one finds the following expression for the expected number of particles:
\begin{equation}
    \frac{\langle N \rangle}{M}=\gamma-\frac{\tau}{2}\gamma^2,
\end{equation}
where
\begin{equation}
    \gamma \tau e^{-\gamma \tau}=\tau e^{-\tau}.
\end{equation}
Then $\gamma$ is given in terms of the Lambert $W$ function
\begin{equation}
    \gamma=-\frac{1}{\tau}W_0(-\tau e^{-\tau}).
\end{equation}

Finally, we arrive at an expression for the typical number density
\begin{equation}
	\frac{\langle N \rangle}{M}=-\frac{1}{\tau}W_0\left(-\tau e^{-\tau} \right)
	-\frac{\tau}{2}\left[\frac{1}{\tau}W_0\left(-\tau e^{-\tau} \right)\right]^2,
\end{equation}
which is consistent with the earlier known result~\cite{lushnikov2005exact}. However, note that our derivation bypasses the need to introduce Mallows-Riordan polynomials, and reproducing the typical number density is another confirmation of the correctness of the replica conjecture.

\begin{figure}
	\includegraphics[width=0.8\columnwidth]{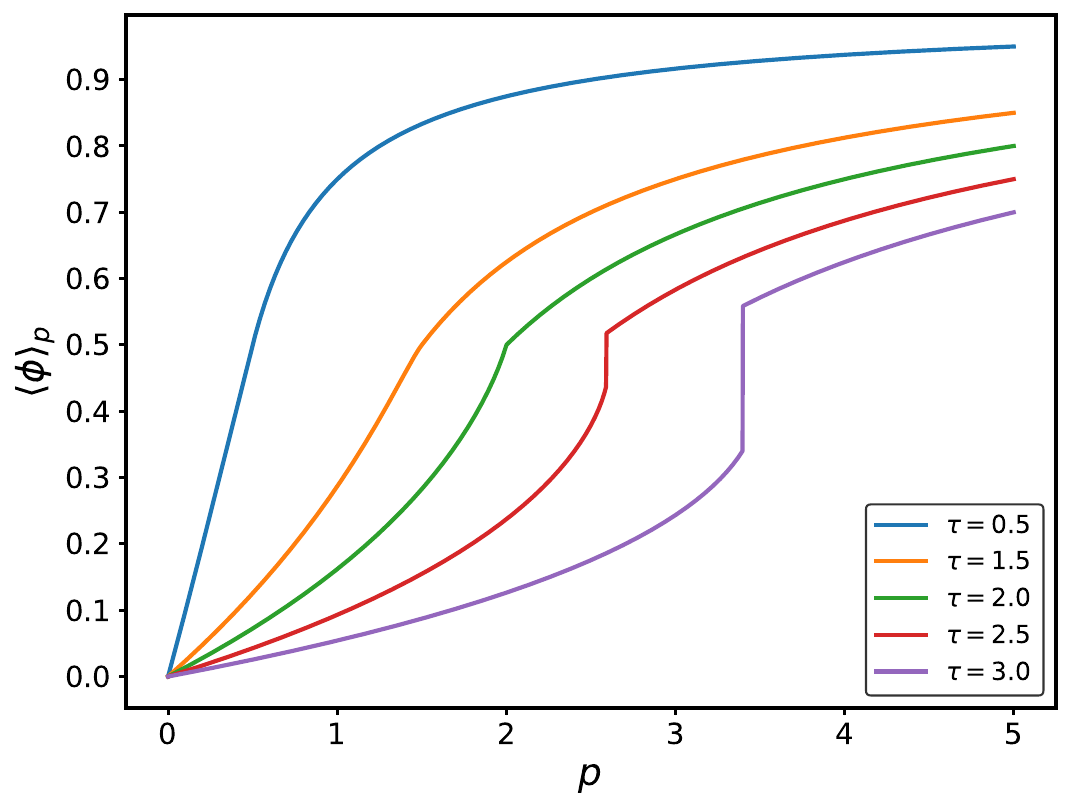}
	\caption{The variation of $\langle \phi \rangle_p$, the mean fraction of particles,  with $p$, for different values of $\tau$. }
	\label{meanphi}
\end{figure}

We also compute the mean number density $\langle \phi \rangle_p$ for a fixed non-zero value of $p$,
\begin{equation}
	\langle \phi \rangle_p=p \frac{d\lambda}{dp}.
\end{equation}
The variation of $\langle \phi \rangle_p$ with $p$ for different values of $\tau$ are shown in Fig.~\ref{meanphi}. For $\tau=0.5, 1.5$ and $2.0$, $\langle \phi \rangle_p$ is a continuous function of $p$, but there is a singular behavior at some $p_c$ as can be seen from the kink-like behavior. On the other hand, for $\tau=2.5$ and $3.0$, $\langle \phi \rangle_p$ has a clear discontinuity at a critical value of $p$.  We now proceed to quantifying the singular behavior in the next section. 

\section{Singular Behavior of Exponential moments}
\label{singular}

In this section, we quantify the singular behavior of the exponential moment.
\subsection{Singularities as a function of $\tau$ for fixed $p$}

We first consider the exponential moment as a function of $\tau$ for fixed $p$.
Our approach is to derive a perturbative expansion for $\Delta F=F^{(c)}(p,\tau,\zeta)-F_0^{(c)}$ around $\tau_c$, where $F_0^{(c)}$ is the classical solution. As established in section~\ref{expmoments}, $\Delta F=0$ when $\tau < \tau_c$. 

We carry out the perturbative expansion separately in three regimes: $p > 2$, $p = 2$, and $p < 2$. For fixed $p$, we introduce a small perturbation around the transition point by setting $\tau = \tau_c + \epsilon$, and then derive the corresponding perturbative expansion for $\zeta$ using Eq.~\eqref{zeta}. For notational simplicity, we define $\nu = (p-2)/(p-1)$, and let $\nu_c$ denote its value at $p = p_c$, where $p_c$ solves Eq.~\eqref{tauc} for fixed $\tau$. Then,
\begin{eqnarray}
	\zeta =
	\begin{cases}
		\displaystyle
		\nu
		+ \frac{\nu^2 }{\nu p - 2 \ln(p-1)} \epsilon +O(\epsilon^2), &p>2,\\
		\displaystyle
		\sqrt{\tfrac{3 \epsilon}{2}}
		- \tfrac{9}{20}\sqrt{\tfrac{3}{2}}\,\epsilon^{3/2}
		+ O(\epsilon^{5/2}), & p=2,\\
		\displaystyle
		-\tfrac{2}{(p-2)p}\,\epsilon
		+ \tfrac{8\,(3+(p-3)p)}{3 (p-2)^3 p^2}\,\epsilon^2 +O(\epsilon^3), & p<2.
	\end{cases}
	\label{zeta-expansions}
\end{eqnarray}

Once the perturbative expansion for $\zeta$ is obtained, we proceed with an analogous expansion for $\Delta F$,
\begin{eqnarray}
	\Delta F =
	\begin{cases}
		\displaystyle
		\frac{ \nu^2 (p-1)\,\epsilon}{2 p} +\\  
		\frac{\nu^3 (p-1)^2\,\epsilon^2}
		{2 p\big(p^2 - 2p + 2\ln(p-1) - 2p\ln(p-1)\big)}
		+ O(\epsilon^3), & p>2, \\
		
		\displaystyle
		\frac{3}{16}\,\epsilon^2
		- \frac{9}{80}\,\epsilon^3
		+ O(\epsilon^{4}), 
		& p=2, \\
		
		\displaystyle
		\frac{2 \epsilon^3}{3 \nu^2 (p \!- \!1) p^3}
		\!-\! \frac{4 (3 \!- \! 3 p +p^2) \epsilon^4}{3 \nu^4 (p-1)^3 p^4} + O(\epsilon^5), \!\!\!
		 & p< 2.  
	\end{cases}
	\label{DeltaF-expansions}
\end{eqnarray}

From the above perturbative expansion, it follows that for $p>2$ the leading non-zero term is proportional to $\epsilon$, indicating that the first derivative of the exponential moment is discontinuous at $\tau_c$. Similarly, for $p=2$, the leading non-zero term is proportional to $\epsilon^2$, implying that the second derivative of the exponential moment is discontinuous at $\tau_c$. Likewise, for $p<2$, the leading non-zero term implies that the third derivative of the exponential moment is discontinuous at $\tau_c$.

\subsection{Singularities as a function of $p$ for fixed $\tau$}
A similar analysis can be carried out for fixed $\tau$ while varying $p$, in which a singularity arises at $p_c$. Introduce a small perturbation at $p_c$ by setting $p=p_c-\epsilon$, and then derive corresponding perturbative expansion for $\zeta$.
\begin{equation}
	\zeta =
	\begin{cases}
		\displaystyle
		\nu_c
		+ \frac{\nu_c^2 \epsilon}{(p_c-1) \left[\nu_c p_c - 2 \ln(p_c-1) \right]}
		+ O(\epsilon^2),\! \!& \tau>2, \\
		
		\displaystyle
		\sqrt{\tfrac{3 \epsilon}{2}} 
		- \tfrac{3}{4} \epsilon
		+ \tfrac{39}{80}\sqrt{\tfrac{3}{2}} \epsilon^{3/2}
		+ O(\epsilon^{2}),  & \tau=2, \\
		
		\displaystyle
		-\frac{2 \epsilon}{(\tau-2)\tau}
		- \frac{4(\tau-3) \epsilon^2}{3(\tau-2)^3\tau}
		+ O(\epsilon^3), & \tau<2.
	\end{cases}
	\label{zeta-pc-expansions}
\end{equation}

From this perturbative expansion, we proceed to construct an  expansion for $\Delta F$,

	\begin{equation}
		\Delta F =
		\begin{cases}
			\displaystyle
			\frac{p_c - 2 - \ln(p_c-1)}{(p_c-1)p_c}\,\epsilon
			+ O(\epsilon^2), 
			&\tau > 2, \\
			
			\displaystyle
			\frac{3}{16}\,\epsilon^2
			- \tfrac{1}{4}\sqrt{\tfrac{3}{2}}\,\epsilon^{5/2}
			+ O(\epsilon^{3}), &
			\tau = 2, \\
			
			\displaystyle
			\frac{2(\tau-1) \epsilon^3}{3(\tau \!- \! 2)^2\tau^3}
			\!+\! \frac{2(2\tau^3 \!- \! 9\tau^2 +12\tau-6) \epsilon^4}{3(\tau-2)^4\tau^4}  \!\!  \!\! &
			\\ + O(\epsilon^5), &
			\tau < 2.
		\end{cases}
		\label{DeltaF-expansions1}
	\end{equation}

Examining the leading non-zero terms, we conclude that for $\tau > 2$, the first derivative of the exponential moment is discontinuous at $p_c$; for $\tau = 2$, the second derivative is discontinuous at $p_c$; and for $\tau < 2$, the third derivative is discontinuous at $p_c$, except at $\tau=1$, where fourth derivative is discontinuous.
\begin{figure}
	\centering
	\includegraphics[width=0.8\columnwidth]{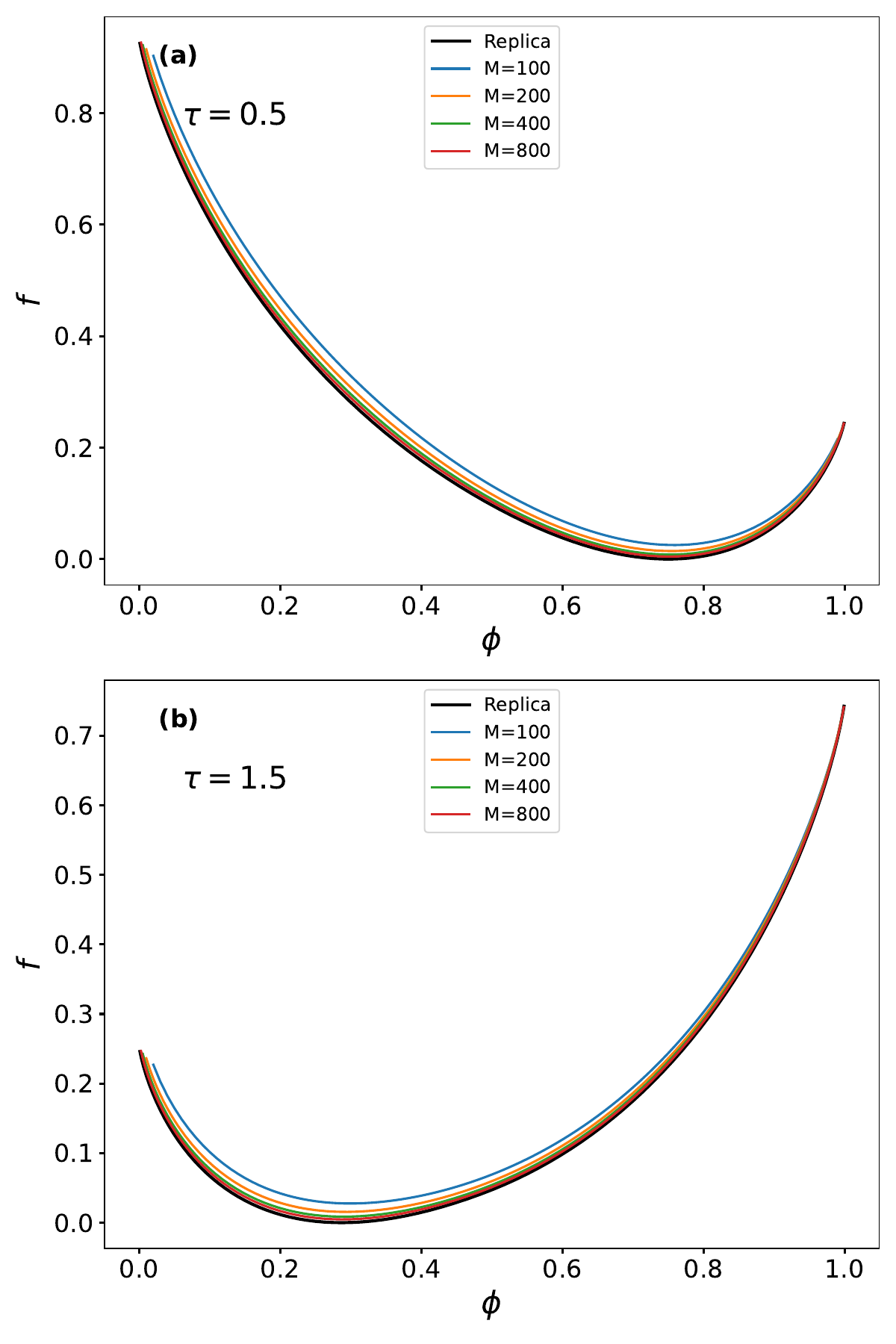}
\caption{Comparison of the LDF obtained from the replica calculation (see Eq.~\eqref{ldf}) with the series expansion (see Eq.~\eqref{prob27}) for finite $M$, for (a) $\tau=0.5$ and (b) $\tau=1.5$.}
	\label{ldfcomparison}
\end{figure}

\section{The Large Deviation Function}
\label{rate}
In this section, we compute the CELDF for product-kernel aggregation. The theory of large deviations primarily addresses two key problems: first, establishing whether a large deviation principle holds for a given random variable, and second, deriving an explicit expression for the corresponding LDF.

A fundamental result in large deviation theory is the G\"artner--Ellis theorem~\cite{gartner1977large,ellis1984large}, which states that if the exponential moment (scaled cumulant generating function) exists and is differentiable, then the random variable satisfies a large deviation principle, and the associated LDF is given by the Legendre--Fenchel transform of the exponential moment. An important property of the Legendre--Fenchel transform is that it always yields a convex function. Therefore, the G\"artner--Ellis theorem cannot be used to calculate non-convex LDF. The breakdown of this theorem is related to the differentiability of the exponential moment $\lambda(p,\tau)$. In Sec.~\ref{singular}, we showed that the exponential moment is non-differentiable for $\tau>2$. If the LDF  $f(\phi,\tau)$ is non-convex, then the Legendre--Fenchel transform of $\lambda(p,\tau)$ does not yield $f(\tau,\phi)$; instead, it yields the convex envelope of $f(\tau,\phi)$~\cite{rockafellar2015convex}.

We now proceed to determine $f(\phi, \tau)$, as defined in Eq.~\eqref{ldf_def}, by taking the Legendre-Fenchel transform of the exponential moment.

For any random variable $X \geq 0$ and any $a > 0$, Markov's inequality implies
\begin{equation}
	P(X \geq a) \leq \frac{\langle e^{\alpha X} \rangle}{e^{\alpha a}} \quad \text{for any $\alpha>0$}.
\end{equation}
Using this inequality, we can write
\begin{equation}
	P(N \geq \phi M) \leq \frac{\langle e^{\alpha N} \rangle}{e^{\alpha M \phi}}.
\end{equation}
To obtain the tightest bound, we minimize over $\alpha$:
\begin{equation}
	P(N \geq \phi M) \leq \inf_{\alpha \geq 0} \langle e^{\alpha N} \rangle e^{-\alpha M \phi}.
\end{equation}
Let $\alpha=\ln p$ (so that $p\geq 1$) noting that $P(N=\phi M)dN\leq P(N \geq \phi M)$. The above estimate shows that
\begin{equation}
	\lim_{M \rightarrow \infty} \frac{1}{M} \ln P(N = \phi M) \leq \inf_{p \geq 1} \left(\frac{1}{M} \ln \langle p^N \rangle -\phi \ln p\right).
\end{equation}
From Eq.~\eqref{replica}, one finds that
\begin{equation}
	-f(\phi, \tau)\leq\inf_{p \geq 1} \left\{\max_{\zeta \geq 0}F^{(c)}(p,\tau,\zeta)-\frac{\tau}{2}-\phi \ln p\right\}.
	\label{ldf}
\end{equation}
The inequality is usually a strict bound unless the LDF is non-convex, and we will take the right hand side to be the CELDF.

We now provide numerical evidence supporting the validity of Eq.~\eqref{ldf}. To do so, we compare the CELDF obtained from the replica calculation for $M\to\infty$, as given by Eq.~\eqref{ldf}, with the LDF derived from the exact expression for finite $M$, given by Eq.~\eqref{prob27}.

Consider Eq.~\eqref{prob27}, in which $P(M,N,t)$ is expressed in terms of the polynomial $\ln\psi_t^{(M)}$ (see Eq.~\eqref{poly}) for finite $M$. For a given value of $M$, the expression $(\ln \psi_t^{(M)}(\mu))^N$ with $N=1,2,\dots,M$ can be expanded as a series in $\mu$, from which the coefficient of $\mu^M$ can be extracted.

\begin{figure}
	\centering
	\includegraphics[width=\columnwidth]{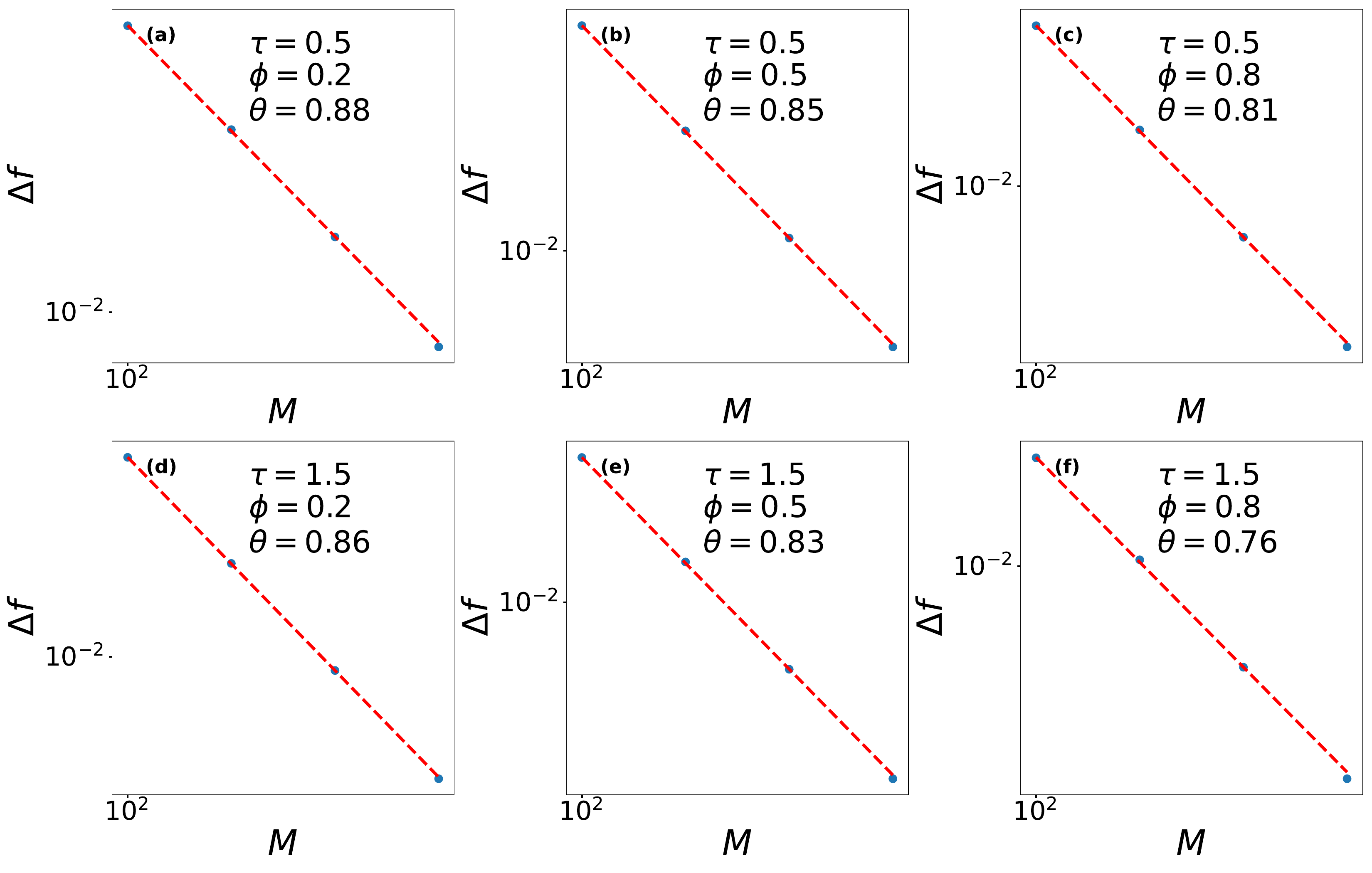}
	\caption{The difference, $\Delta f$, between the LDF obtained from the series expansion for finite $M$, and the LDF from the replica calculation for infinite $M$. The straight lines are power-laws $M^{-\theta}$ with exponent $\theta$ as shown in figure.  The data are  for (a) $\tau=0.5$, $\phi=0.2$, (b) $\tau=0.5$, $\phi=0.5$, (c) $\tau=0.5$, $\phi=0.8$, (d) $\tau=1.5$, $\phi=0.2$,(e) $\tau=1.5$, $\phi=0.5$, and (f) $\tau=1.5$, $\phi=0.8$.}
	\label{ldfconvergence}
\end{figure}

The LDF for different values of $M$ are compared with the CELDF from the replica calculation (Eq.~\eqref{ldf}) in Fig.~\ref{ldfcomparison}(a) [$\tau=0.5$] and Fig.~\ref{ldfcomparison}(b) [$\tau=1.5$]. In computing the CELDF using Eq.~\eqref{ldf}, we determine the optimal value $\zeta^*$ that maximizes Eq.~\eqref{ldf}.  Even though the curves are close to each other, discrepancies are visible in Fig.~\ref{ldfcomparison}(a) and (b). We show that these differences are finite-size corrections and vanish in the limit $M\to\infty$ by demonstrating that the difference $\Delta f$ exhibits a power-law dependence on $M$ for various values of $\phi$, $i.e$,
\begin{equation}
	\Delta f(\phi)=\frac{g(\phi)}{M^\theta}.
	\label{powerlaw}
\end{equation}
We compute the exponent $\theta$ for three representative values: $\phi=0.2$, $\phi=0.5$, and $\phi=0.8$ (see Fig.~\ref{ldfconvergence}(a)–(c) for $\tau=0.5$ and Fig.~\ref{ldfconvergence}(d)–(f) for $\tau=1.5$). The observed variation of the scaling exponent $\theta$ with $\phi$ in Fig.~\ref{ldfconvergence} is likely a numerical artifact arising from the limited system sizes ($M = 100, 200, 400, 800$) available for the power-law fit, rather than a genuine physical effect, and we expect $\theta$ to be independent of $\phi$. 

\begin{figure}
	\includegraphics[width=\columnwidth]{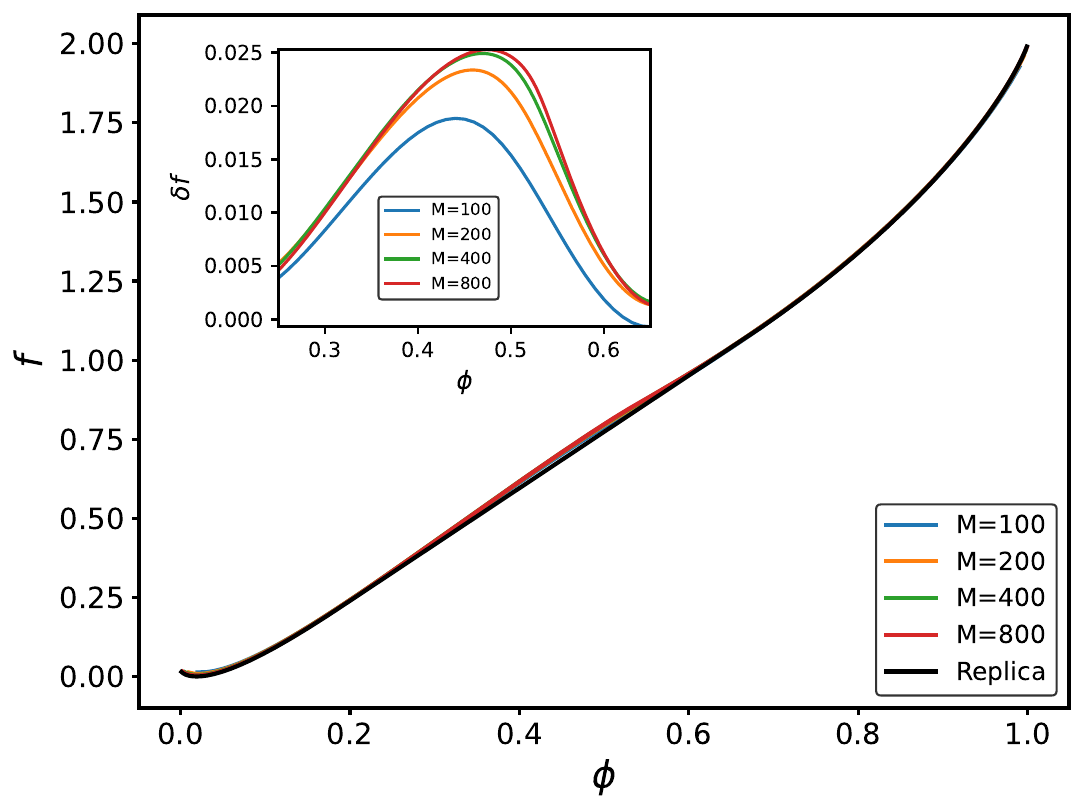}
	\caption{Comparison between the CELDF derived from the replica calculation (Eq.~\eqref{ldf}) and the LDF obtained via series expansion for finite $M$ (Eq.~\eqref{prob27}) when $\tau=4$. Inset: Difference between the LDF for finite $M$ and the CELDF, near the non-convex region.
	}
	\label{ratefunction}
\end{figure}

For $\tau>2$, the exact LDF obtained from Eq.~\eqref{prob27} is non-convex since the exponential moment is not differentiable. Consequently Eq.~\eqref{ldf} yields the convex envelope of the exact LDF. In Fig.~\ref{ratefunction} we compared the LDF for finite $M=100,200,400,800$ using Eq.~\eqref{prob27} with the CELDF from the replica result given by Eq.~\eqref{ldf}. The inset shows the difference, $\delta f$, between the LDF for finite $M$ and the CELDF from the replica calculation near the non-convex region. It is evident that this difference grows with increasing mass.

The classical solution, which corresponds to $\zeta=0$, is given by
\begin{equation}
	f(\tau,\phi)=\sup_{p \geq 1} \left\{-\frac{\tau}{2p}-\ln p+\frac{\tau}{2}+\phi \ln p\right\}.
\end{equation}
Minimization of above expression leads to the optimal value of $p$,
\begin{equation}
	p^*=\frac{\tau}{2(1-\phi)}.
	\label{classp}
\end{equation}
Therefore, the classical LDF is given by
\begin{equation}
	f_{cl}(\tau,\phi)=\frac{\tau}{2}-(1-\phi)\left[1-\ln \left(\frac{2(1-\phi)}{\tau}\right)\right].
	\label{classldf}
\end{equation}

\section{Singular Behavior of CELDF \label{ldfsingular}}

\begin{figure}
	\includegraphics[width=\columnwidth]{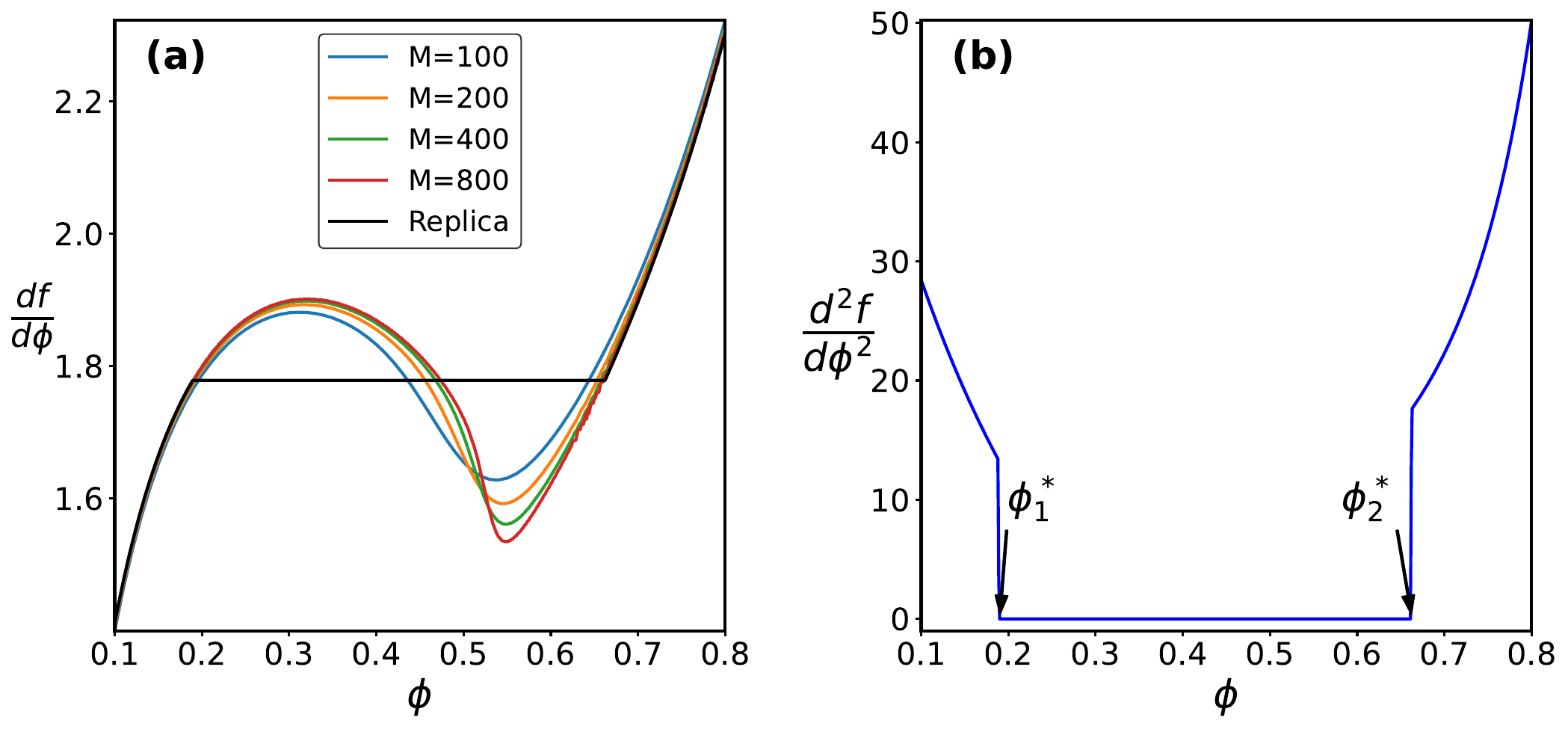}
		\caption{(a) Comparison of the first derivative of the CELDF from the replica calculation (Eq.~\eqref{ldf}) with that obtained from the series expansion for finite $M$ (Eq.~\eqref{prob27}) at $\tau=4$. (b) The second derivative of the CELDF obtained from the replica calculation for $\tau=4$. The second derivative shows discontinuities at two points, denoted by $\phi_1^*$ and $\phi_2^*$.}
	\label{ldfderivatives}
\end{figure}

In this section, we examine the singular behavior of the CELDF, resulting in the diagram. From Eq.~\eqref{ldf}, we have
\begin{equation}
	\frac{df}{d\phi}=\ln p^*,
\end{equation}
where $p^*$ is the optimal value at which Eq.~\eqref{ldf} is minimized. From Fig.~\ref{ldfderivatives}(a), it is evident that the interval of $\phi$ over which $df/d\phi$ is constant corresponds to the non-convex region of the LDF. Moreover, $df/d\phi$ for finite $M$, obtained from Eq.~\eqref{prob27}, deviates increasingly from the convex envelope as $M$ increases. This indicates that the non-convexity is not a finite-size effect. 

Section~\ref{rate} established that for $\tau>2$ the replica calculation produces the convex envelope of the exact LDF. Accordingly, the second derivative $d^2 f/d\phi^2$, of the CELDF exhibits discontinuities at the two end points of the convex envelope (Fig.~\ref{ldfderivatives}(b)). We denote the first and second points at which these discontinuities occur by $\phi_1^*$ and $\phi_2^*$, respectively. In the following, we compute these points and discuss their significance.

\subsection{Calculating $\phi_1^*$ and $\phi_2^*$}
While numerically evaluating Eq.~\eqref{ldf}, we compute the optimal value of $p$ that minimizes Eq.~\eqref{ldf}, denoted by $p^*$, along with the corresponding value of $\zeta$, denoted by $\zeta^*$.
\begin{figure}
	\includegraphics[width=\columnwidth]{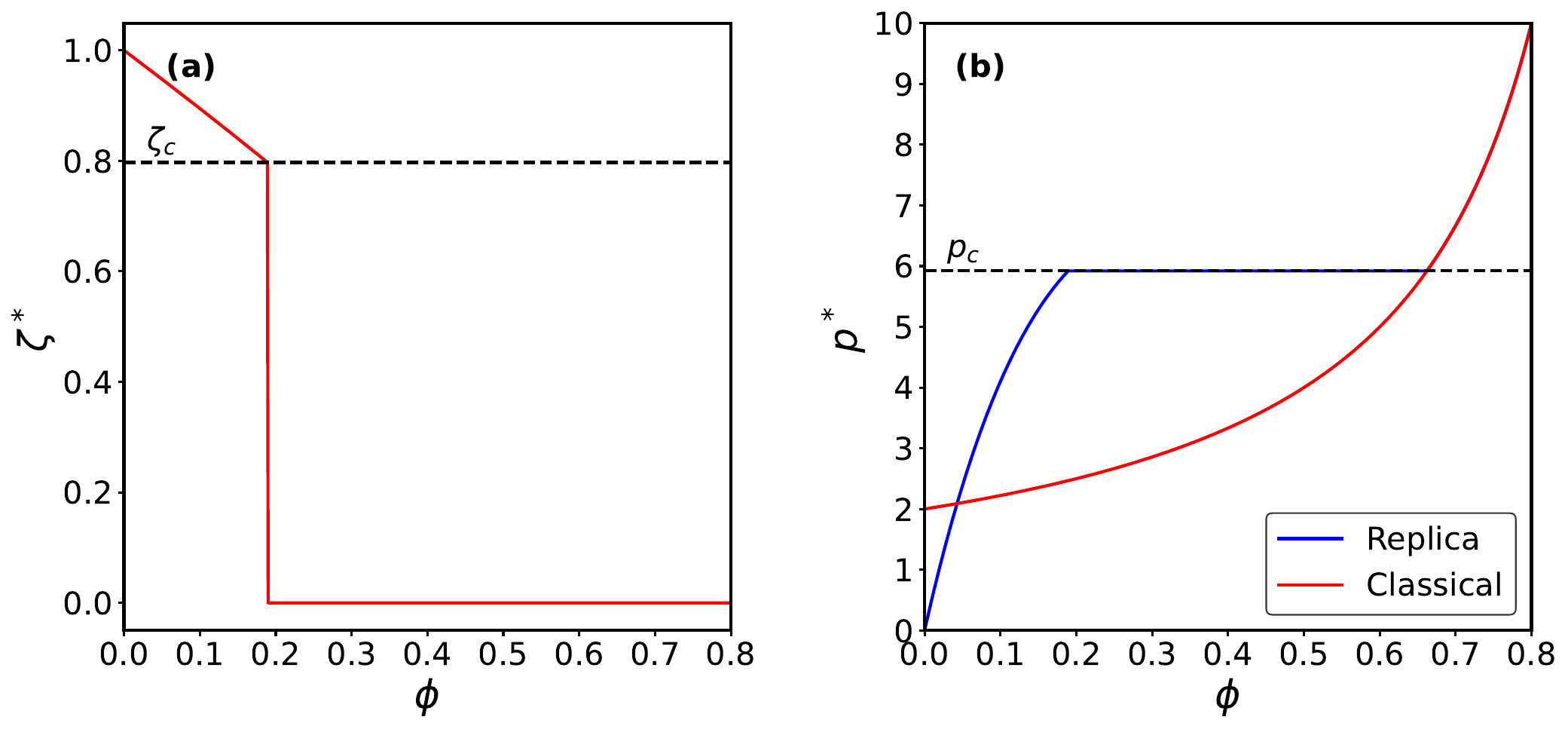}
 		\caption{(a) Variation of $\zeta^*$ with $\phi$ for $\tau=4$. At $\phi_1^*$, $\zeta^*$ exhibits a discontinuous transition from a non-zero value to zero, and beyond $\phi_1^*$ it remains zero. (b) Variation of $p^*$ with $\phi$ for $\tau=4$. Beyond $\phi_2^*$, $p^*$ coincides with the expression in Eq.~\eqref{classp}, corresponding to $\zeta=0$.}
	\label{zetap}
\end{figure}
From Fig.~\ref{zetap}, the significance of the points $\phi_1^*$ and $\phi_2^*$ is evident. Up to $\phi_1^*$, the replica solution yields non-zero values of $\zeta^*$; beyond this point, $\zeta^*$ becomes zero. The point $\phi_2^*$ corresponds to the onset of the classical solution given by Eq.~\eqref{classldf}.

Between $\phi_1^*$ and $\phi_2^*$, $p^*$ takes a constant value, denoted by $p_c$ in Fig.~\ref{zetap}(b), which is the solution of Eq.~\eqref{tauc} for a given $\tau$. The corresponding value $\zeta_c$, marked in Fig.~\ref{zetap}(a), is given by Eq.~\eqref{zetac}. 

Applying the minimization condition to Eq.~\eqref{ldf} at $\phi_1^*$ leads to
\begin{equation}
	\phi_1^*=p_c \left[\frac{\partial F^{(c)}}{\partial p}+\frac{\partial F^{(c)}}{\partial \zeta}\frac{\partial F^{(c)}}{\partial p}\right]_{p_c,\zeta_c}.
\end{equation}
Since $p_c$ and $\zeta_c$ satisfy the relations \eqref{tauc} and \eqref{zetac}, respectively,
\begin{equation}
	\phi_1^*=\frac{p_c(p_c-2)-\ln (p_c-1)}{p_c(p_c-1)(p_c-2)},\quad\text{$\tau > 2$}.
\end{equation}
For $\tau < 2$, $p_c=\tau$ and $\zeta_c=0$, hence
\begin{equation}
		\phi_1^*=\frac{1}{2},\quad\text{$\tau \leq 2$}.
\end{equation}
Using Eq.~\eqref{classp} at $\phi_2^*$,
\begin{equation}
	p(\phi_2^*)=p_c,
\end{equation}
leads to 
\begin{equation}
	\phi_2^*=1-\frac{\tau}{2p_c}, \quad\text{$\tau >2$}.
\end{equation}
For $\tau<2$, $p_c=\tau$ and
\begin{equation}
	\phi_2^*=\frac{1}{2} \quad\text{$\tau\leq 2$}.
\end{equation}

\begin{figure}
	\includegraphics[width=\columnwidth]{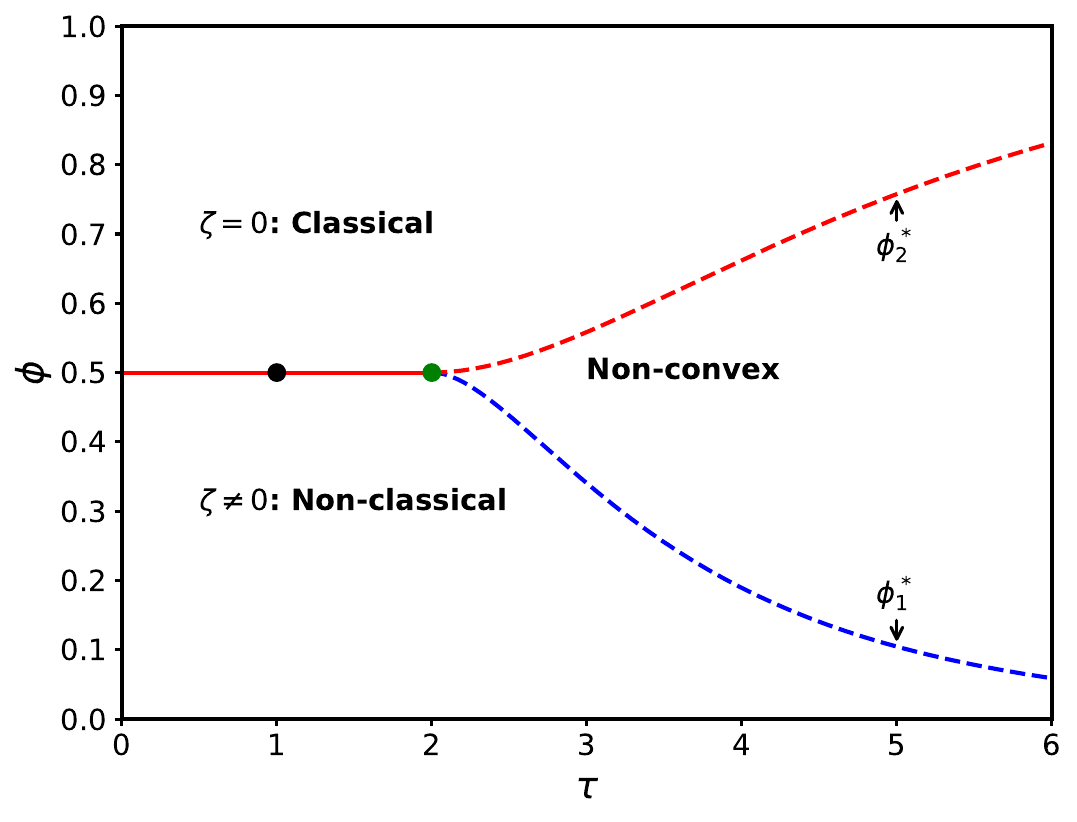}
	\caption{The  phase diagram for product-kernel aggregation in the $\phi$--$\tau$ plane. The dashed lines represent a first-order transition while the solid line denotes a second-order transition. The green circle at $(1/2, 2)$ is a tricritical point, while the black circle at  $(1/2, 1)$, (corresponding to typical evolution) has a different singular behavior (see text).}
	\label{phasediagram}
\end{figure}

Having calculated the two transition points, in Fig.~\ref{phasediagram} we construct a complete phase diagram for product-kernel aggregation in the $\phi$--$\tau$ plane. The phase diagram can be divided into three distinct regions. The classical region ($\zeta = 0$) is characterized by the absence of a gel, whereas the non-classical region ($\zeta \neq 0$) exhibits gel formation. The region in which the LDF becomes non-convex corresponds to phase coexistence.
\subsection{Singular behavior for $\tau\leq2$ }
In Sec.~\ref{ldfsingular}, we observed that for $\tau > 2$ the second derivative of the CELDF is discontinuous. In this section, we investigate the singular behavior of the CELDF for $\tau \leq 2$.

For $\tau \leq 2$, the singularity occurs at $\phi_c = 1/2$. For fixed $\tau$, we have already derived a perturbative expansion for $\zeta$ (see Eq.~\eqref{zeta-pc-expansions}). We introduce a small perturbation around $\phi_c$ by setting $\phi=1/2-\delta$, and obtain a corresponding expansion for $\Delta f=f(\tau,\phi)-f_{cl}(\tau,\phi)$, where $f(\tau,\phi)$ and $f_{cl}(\tau,\phi)$ are given by Eqs.~\eqref{ldf} and \eqref{classldf}, respectively.
\begin{eqnarray}
	\Delta f =
	\begin{cases}
		\displaystyle
		\frac{16(\tau-1) \delta^3}{3(\tau-2)^2}
		- \frac{32\tau^3 \delta^4}{3(\tau-2)^4}
		+ O(\delta^5), \!\! \! \!
		& \tau<2,\ \\
		
		\displaystyle
		- \frac{32}{3}\,\delta^4+O(\delta^5),
		& \tau=1, \\
		
		\displaystyle
		\frac{3}{4}\,\delta^2
		- \tfrac{1}{4}\sqrt{\tfrac{3}{2}}\,\delta^{5/2}
		+ O(\delta^{3}),
		& \tau=2.
	\end{cases}
	\label{Deltaf-expansions}
\end{eqnarray}
For $\phi>\phi_c$, $\Delta f=0$.

Examining the leading non-zero terms, we find that for $\tau = 2$ the second derivative of the CELDF is discontinuous. For $\tau < 2$, the discontinuity occurs in the third derivative, except at $\tau = 1$, where the discontinuity appears in the fourth derivative.
\section{Perturbative Expansion of the CELDF for small $\phi$}
\label{perturbation}
\begin{figure}
	\includegraphics[width=0.8\columnwidth]{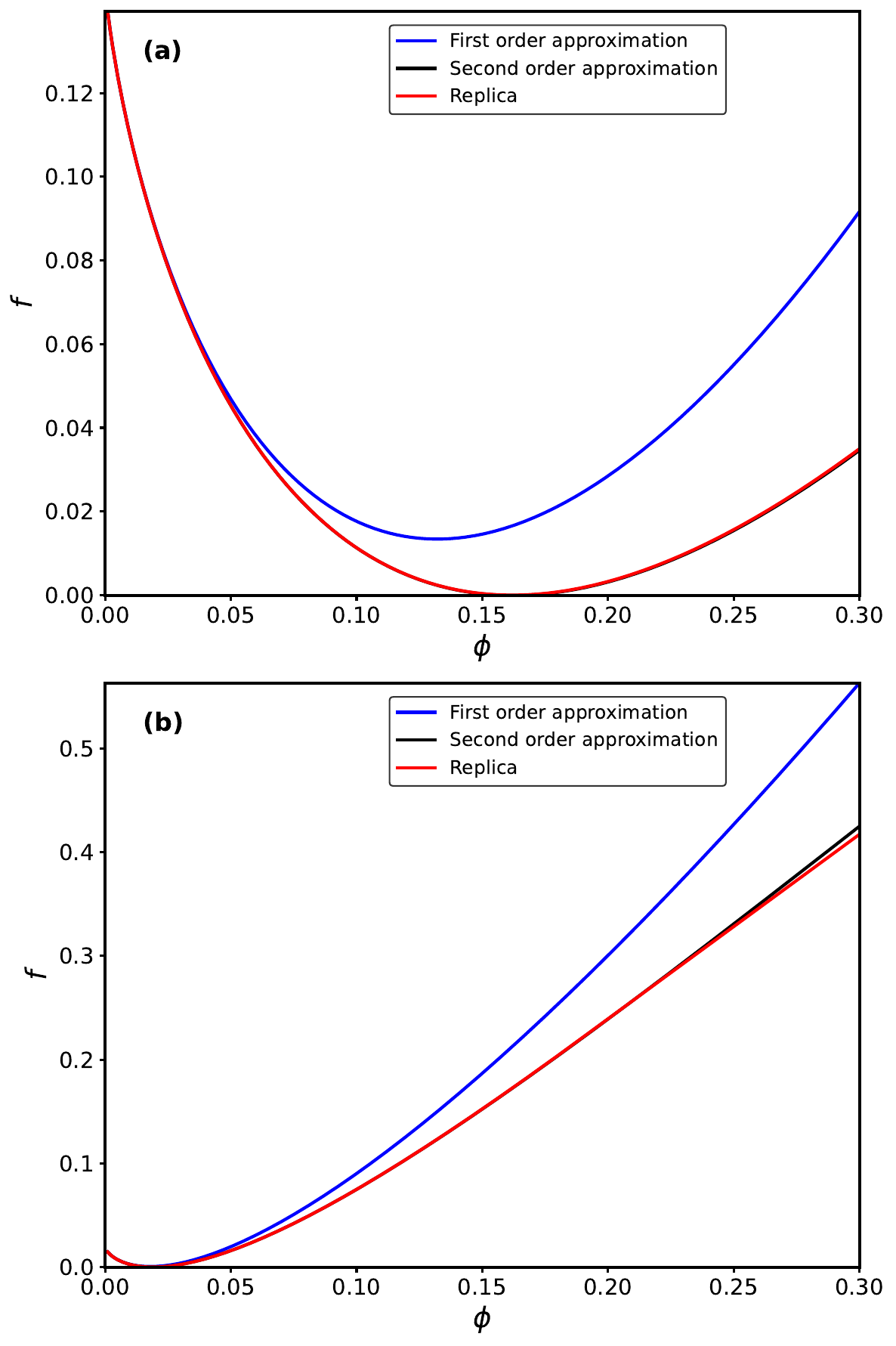}
	\caption{Comparison of the perturbative expansion for the LDF given by Eq.~\eqref{pertbldf} with the exact result (Eq.~\eqref{ldf}), for (a) $\tau=2$ and (b) $\tau=4$.}
	\label{ldfperturb}
\end{figure}

The goal of this section is to derive a perturbative expansion for the CELDF obtained from the replica calculation in the limit of small $\phi$. From Fig.~\ref{zetap}(b), we observe that as $\phi \to 0$, $p$ also tends to zero. To obtain a perturbative expansion of $p$ in terms of $\phi$, we first construct a perturbative expansion of $\zeta$ in powers of $p$. The relation satisfied by $\zeta$ is
\begin{equation}
	\zeta=\frac{e^{\tau \zeta}-1}{p-1+e^{\tau \zeta}}.
\end{equation}
A perturbative expansion for $\zeta$ in terms of $p$ gives
\begin{equation}
	\zeta=1-\frac{p}{(e^\tau-1)}+\frac{e^\tau(1-\tau)-1}{(e^\tau-1)^3}p^2+O(p^3),
	\label{zetaseries}
\end{equation}
where the expansion is truncated at second order. Using Eq.~\eqref{zetaseries}, we obtain the optimal value of $p$ in Eq.~\eqref{ldf} by solving
\begin{equation}
	\frac{dF^{(c)}}{dp}-\frac{\phi}{p}=0,
\end{equation}
which yields the following expansion for $p$:
\begin{equation}
	\begin{split}
		&p=\left[\frac{2 \left(e^{\tau } -1 \right)^2}{-2+2 e^{\tau }-\tau }\right]\phi\\
		&-\left[\frac{8 \left(e^{\tau} - 1 \right)^2 \left(1-e^{2
				\tau } (\tau-1 )+\tau +e^{\tau } \left(\tau^2 - 2 \right)\right)}{\left(-2+2 e^{\tau }-\tau \right)^2 \left(2-2 e^{\tau }+\tau \right)}\right]\phi^2\\&+O(\phi^3).
	\end{split}
	\label{pertb}
\end{equation}
Substituting Eq.~\eqref{pertb} back into Eq.~\eqref{ldf}, we obtain a perturbative expansion for the CELDF:
\begin{equation}
	\begin{split}
		\raisetag{30pt}
		&f(\tau,\phi)=-\ln\left[1-e^{-\tau }\right] + \ln\left[\frac{2 \left(-1+e^{\tau }\right)^2 \phi }{(-2+2 e^{\tau }-\tau) e } \right] 
		\phi \\
		&-\left[\frac{2 \left(-1+2 e^{\tau }-e^{2 \tau }-\tau +e^{2 \tau } \tau -e^{\tau } \tau ^2\right) }{\left(-2+2 e^{\tau }-\tau \right)^2}\right] \phi^2\\&+\phi \ln \phi+O(\phi^3).
	\end{split}
	\label{pertbldf}
\end{equation}
In Fig.~\ref{ldfperturb}, we compare Eq.~\eqref{pertbldf} with the replica result for $\tau=2$ and $\tau=4$, and observe good agreement for small $\phi$.

\section{Summary and Discussion}
\label{summary}

In summary, we presented an exact and comprehensive analysis of rare fluctuations in product--kernel cluster--cluster aggregation. We focused on  the quantity $P(M, N, t)$, the  probability that $N$ particles remain at time $t$ when starting with $M$ particles of mass $1$.  Starting directly from the master equation, we derived an exact integral representation for $P(M,N,t)$, valid for arbitrary finite values of $M$, $N$, and $t$. This representation provides a controlled starting point for studying large deviations beyond the typical evolution described by the Smoluchowski equation.

From the exact expression for $P(M,N,t)$, we obtained closed-form results for the exponential moments $\langle p^{N} \rangle$ for integer $p$. Using a replica conjecture---supported by detailed finite-$M$ numerical scaling---we extended the exponential moments to real $p \geq 0$. A careful asymptotic analysis revealed singular behavior in these moments, signaling the presence of phase transitions in the associated large-deviation function.  This allowed us to compute the exact convex envelope of the large-deviation function  for all values of the scaled particle fraction $\phi = N/M$ and scaled time $\tau = tM$.

The singular structure of the CELDF yields a rich phase diagram for product--kernel aggregation. We identified regions in which the underlying large-deviation function is non-convex, corresponding to coexistence between distinct fluctuation mechanisms, as well as regions where the transition is continuous. These regimes are separated by analytically determined phase boundaries that meet at a tricritical point.

We summarize the singular behavior of the exponential moment  and CELDF.  For the exponential moment with fixed $p$ and varying $\tau$, the first derivative is discontinuous for $p > 2$, the second derivative for $p = 2$, and the third derivative for $p < 2$. Similarly, for fixed $\tau$ and varying $p$, the first, second, and third derivatives are discontinuous for $\tau > 2$, $\tau = 2$, and $\tau < 2$, respectively.  For the CELDF,  when  $\tau\geq 2$, the second derivative of the CELDF is discontinuous, while for $\tau<2$ the discontinuity occurs in the third derivative, except at $\tau=1$, where it appears in the fourth derivative. 

Our approach complements and extends earlier work based on mappings to Erd\H{o}s--R\'enyi random graphs and mean-field Potts models. In contrast to those approaches, the present framework does not rely on non-bijective mappings or assumptions about analytic continuation, and remains entirely within the aggregation dynamics itself. As a result, it yields exact finite-$M$ expressions, controlled asymptotics, and a transparent route to generalization.

The derivation of a formula for $P(M,N,t)$ allows us to obtain the non-convex part of the LDF. We note that the Legendre transform of the exponential moment can only give the CELDF. Also, in biased Monte Carlo simulations, only the CELDF can be obtained. In this paper, we have not addressed the asymptotic limit of the non-convex regime. To do so, it would involve solving the Euler Lagrange equations for the mass distribution along the optimal trajectories. This is a promising area for future study.

Several extensions suggest themselves naturally. The Doi-Peliti-Zeldovich formalism that we use in the paper is general and can be adapted to computing other large deviation questions of interest in product kernel aggregation (as well as other reaction-diffusion systems). For instance, one could probe the role of initial conditions, going beyond the monodispersed state considered in this paper. Other quantities that can be studied are $P(M,N_k,t)$, the probability for $N_k$ particles of mass $k$ to remain at time $t$. Preliminary calculations show that the LDF for $P(M,N_k,t)$, has non-convex behavior which can be fully characterized. The methods can also be possibly applied to recent studies of gelation in related problems~\cite{krapivsky2024gelation,dyachenko2023finite}. In low dimensions, spatial fluctuations become dominant and require analytical or numerical treatments beyond mean-field theory~\cite{spouge1988exact,kang1984fluctuation,krishnamurthy2002kang,krishnamurthy2003persistence,ravindran2025finite}.  Generalization of the methods in this paper to study rare fluctuations in these  systems is an important problem.

%

\appendix

\section{Expressing $P(M,N,t)$ in terms of annihilation and creation operators \label{appen}}

The master equation, Eq.~\eqref{master}, can be written in the form of Schr\"{o}dinger equation by introducing creation and annihilation operators. The number of particles of mass $i$, $N_i$, is denoted as the eigenvalue of a number operator $\hat{N}_i=a_i^\dagger a_i$ acting on a state $\vert \vec{N} \rangle=\vert N_1,N_2,...N_M \rangle$,
\begin{equation}
	\hat{N}_i \vert \vec{N} \rangle=N_i  \vert \vec{N} \rangle.
\end{equation}
The annihilation and creation operators obey the canonical commutation relation $[a_m,a_n^\dagger]=\delta_{m,n}$ and have following properties
\begin{align}
	a_i \vert \vec{N} \rangle &= N_i \vert \vec{N}-I_i \rangle, \\
	a_i^\dagger \vert \vec{N} \rangle &= \vert \vec{N}+I_i \rangle, 
\end{align}
where $I_i$ denotes the change in $\vert \vec{N}\rangle$ through the increase or decrease of the number of clusters of mass $i$ by $1$.

Consider the state vector $\vert \psi(t) \rangle$:
\begin{equation}
	\vert \psi(t) \rangle=\sum_{\vec{N}}P(\vec{N},t) \vert \vec{N} \rangle.
	\label{state}
\end{equation}
Multiplying both sides of the master equation Eq.~\eqref{master} by $\vert \vec{N} \rangle$ and summing over all configurations $\vec{N}$ we obtain an evolution equation for $\vert \psi(t) \rangle$, which using the properties of annihilation and creation operators can be written as
\begin{equation}
	\frac{d \vert \psi(t) \rangle}{dt}=-H(\{a_i^\dagger,a_i\})\vert \psi(t) \rangle,
	\label{schr}
\end{equation}
where the Hamiltonian $H$ is
\begin{equation}
	H(\{a_i^\dagger,a_i\})=-\frac{1}{2}\sum_i \sum_j m_im_j\left(a_{i+j}^\dagger-a_i^\dagger a_j^\dagger\right)a_i a_j.
\end{equation}
The solution of Eq~\eqref{schr} is 
\begin{equation}
	\vert \psi(t) \rangle =e^{-Ht}\vert \psi(0) \rangle.
\end{equation}

At time $t=0$, there are $M$ particles of mass $1$, $i.e$, $\vert \psi(t) \rangle =e^{-Ht}a_1^{\dagger M} \vert 0 \rangle$. Multiply both side of Eq.~\eqref{state} by $\langle \vec{N} \vert$ and using the result $\langle \vec{M} \vert \vec{N} \rangle=\frac{1}{\vec{N}!}\delta_{\vec{M},\vec{N}}$ we obtain
\begin{equation}
	P(\vec{N},t)=\frac{\langle \vec{N} \vert \psi(t) \rangle}{\vec{N}!},
\end{equation}
where $\vec{N}!=N_1! N_2! ...N_M!$.
Thus, $P(M,N,t)$ can be written as
\begin{equation}
	P(M,N,t)=\sum_{\vec{N}}\langle 0 \vert \left(\prod_{i=1}^{M}\frac{a_i^{N_i}}{N_i!}\right)\delta \left(\sum_{\vec{N}}N_i-N\right) \vert \psi(t) \rangle.
\end{equation}
Multiplying and dividing by $N!$, and using multinomial theorem we obtain
\begin{equation}
	P(M,N,t)=\frac{1}{N!} \langle 0 \vert \left(\sum_{k=1}^{M}a_k\right)^N e^{-Ht}a_1^{\dagger M} \vert 0 \rangle.
	\label{std}
\end{equation}

\section{Methods of numerical evaluation}
\label{numerical_app}
In this appendix, we provide a detailed account of the numerical evaluation of the LDF for finite $M$.

{\it Evaluation of $P(M,N,t)$:} From Sec.~\ref{intrep}, we have the exact integral representation
\begin{equation}
	P(M,N,t)=\frac{ e^{-\frac{M^2t}{2}}M!}{N!} \oint \frac{d\mu}{2\pi i \mu^{M+1}}\left[\ln \psi_t^{(M)}(\mu)\right]^N,
	\label{prob-app}
\end{equation}
where
\begin{equation}
	\ln \psi_t(\mu)=\sum_{n=1}^{\infty}\left(e^t -1\right)^{n-1}e^{\frac{nt}{2}}F_{n-1}\left(e^t\right)\frac{\mu^n}{n!}.
	\label{series_app}
\end{equation}

For fixed $M$ and $\tau$, the Mallows--Riordan polynomials $F_n(e^t)$ are generated using the recursion relation Eq.~\eqref{recursion}. Once these polynomials are known, the quantity $\left[\ln \psi_t^{(M)}(\mu)\right]^N$ is expanded as a power series in $\mu$ for a given $N$. The coefficient of $\mu^M$ in this expansion is then extracted, yielding $P(M,N,t)$ via Eq.~\eqref{prob-app}. Since only the coefficient of $\mu^M$ contributes to the contour integral, the summation in Eq.~\eqref{series_app} can be truncated at $n=M$ without loss of accuracy. During the evaluation, the coefficients of the polynomial become extremely large, leading to numerical errors. To circumvent this issue, we convert all coefficients to logarithms and perform polynomial multiplication in the log scale.

{\it Evaluation of $\langle p^N \rangle$:} In Sec.~\ref{expmoments}, we derived an exact expression for the exponential moments:
\begin{equation}
	\langle p^N \rangle = e^{-\frac{M^2t}{2}} M! \oint \frac{d\mu}{2\pi i \mu^{M+1}} \left(\psi_t^{(M)}(\mu)\right)^p,
	\label{expmoment_app}
\end{equation}
where
\begin{equation}
	\psi_t^{(M)}(\mu) = \sum_{n=0}^{M} \frac{\mu^n}{n!} e^{\frac{n^2 t}{2}}.
	\label{psi_app}
\end{equation}

For integer $p$, the evaluation is straightforward: expand $\left(\psi_t^{(M)}(\mu)\right)^p$ as a power series in $\mu$, extract the coefficient of $\mu^M$, and obtain $\langle p^N \rangle$ from the contour integral. However, we are interested in real $p>0$. For such values, rather than taking a fractional power directly, we rewrite
\begin{equation}
	\left(\psi_t^{(M)}(\mu)\right)^p = \exp\bigl(p\,h(\mu)\bigr),
	\label{psi_exp_app}
\end{equation}
where $h(\mu)=\ln \psi_t(\mu)$ is given by Eq.~\eqref{series_app}. Expanding the exponential then yields
\begin{equation}
	\left(\psi_t^{(M)}(\mu)\right)^p = 1 + p\,h(\mu) + \frac{p^2 h^2(\mu)}{2!} + \cdots + \frac{p^M h^M(\mu)}{M!}.
	\label{frac_app}
\end{equation}
Since only the coefficient of $\mu^M$ contributes to the contour integral in Eq.~\eqref{expmoment_app}, the expansion can be truncated at $p^M$. For fixed $M$, $p$, and $\tau$, each term in Eq.~\eqref{frac_app} is evaluated, the coefficient of $\mu^M$ is extracted, and $\langle p^N \rangle$ is obtained.


\begin{thebibliography}{50}%
	\makeatletter
	\providecommand \@ifxundefined [1]{%
		\@ifx{#1\undefined}
	}%
	\providecommand \@ifnum [1]{%
		\ifnum #1\expandafter \@firstoftwo
		\else \expandafter \@secondoftwo
		\fi
	}%
	\providecommand \@ifx [1]{%
		\ifx #1\expandafter \@firstoftwo
		\else \expandafter \@secondoftwo
		\fi
	}%
	\providecommand \natexlab [1]{#1}%
	\providecommand \enquote  [1]{``#1''}%
	\providecommand \bibnamefont  [1]{#1}%
	\providecommand \bibfnamefont [1]{#1}%
	\providecommand \citenamefont [1]{#1}%
	\providecommand \href@noop [0]{\@secondoftwo}%
	\providecommand \href [0]{\begingroup \@sanitize@url \@href}%
	\providecommand \@href[1]{\@@startlink{#1}\@@href}%
	\providecommand \@@href[1]{\endgroup#1\@@endlink}%
	\providecommand \@sanitize@url [0]{\catcode `\\12\catcode `\$12\catcode
		`\&12\catcode `\#12\catcode `\^12\catcode `\_12\catcode `\%12\relax}%
	\providecommand \@@startlink[1]{}%
	\providecommand \@@endlink[0]{}%
	\providecommand \url  [0]{\begingroup\@sanitize@url \@url }%
	\providecommand \@url [1]{\endgroup\@href {#1}{\urlprefix }}%
	\providecommand \urlprefix  [0]{URL }%
	\providecommand \Eprint [0]{\href }%
	\providecommand \doibase [0]{https://doi.org/}%
	\providecommand \selectlanguage [0]{\@gobble}%
	\providecommand \bibinfo  [0]{\@secondoftwo}%
	\providecommand \bibfield  [0]{\@secondoftwo}%
	\providecommand \translation [1]{[#1]}%
	\providecommand \BibitemOpen [0]{}%
	\providecommand \bibitemStop [0]{}%
	\providecommand \bibitemNoStop [0]{.\EOS\space}%
	\providecommand \EOS [0]{\spacefactor3000\relax}%
	\providecommand \BibitemShut  [1]{\csname bibitem#1\endcsname}%
	\let\auto@bib@innerbib\@empty
	\bibitem [{\citenamefont {Guria}\ \emph {et~al.}(2009)\citenamefont {Guria},
		\citenamefont {Herrero},\ and\ \citenamefont
		{Zlobina}}]{guria2009mathematical}%
	\BibitemOpen
	\bibfield  {author} {\bibinfo {author} {\bibfnamefont {G.~T.}\ \bibnamefont
			{Guria}}, \bibinfo {author} {\bibfnamefont {M.~A.}\ \bibnamefont {Herrero}},\
		and\ \bibinfo {author} {\bibfnamefont {K.~E.}\ \bibnamefont {Zlobina}},\
	}\bibfield  {title} {\bibinfo {title} {A mathematical model of blood
			coagulation induced by activation sources},\ }\href@noop {} {\bibfield
		{journal} {\bibinfo  {journal} {Discrete and Continuous Dynamical Systems}\
		}\textbf {\bibinfo {volume} {25}},\ \bibinfo {pages} {175} (\bibinfo {year}
		{2009})}\BibitemShut {NoStop}%
	\bibitem [{\citenamefont {Falkovich}\ \emph {et~al.}(2002)\citenamefont
		{Falkovich}, \citenamefont {Fouxon},\ and\ \citenamefont
		{Stepanov}}]{falkovich2002acceleration}%
	\BibitemOpen
	\bibfield  {author} {\bibinfo {author} {\bibfnamefont {G.}~\bibnamefont
			{Falkovich}}, \bibinfo {author} {\bibfnamefont {A.}~\bibnamefont {Fouxon}},\
		and\ \bibinfo {author} {\bibfnamefont {M.}~\bibnamefont {Stepanov}},\
	}\bibfield  {title} {\bibinfo {title} {Acceleration of rain initiation by
			cloud turbulence},\ }\href@noop {} {\bibfield  {journal} {\bibinfo  {journal}
			{Nature}\ }\textbf {\bibinfo {volume} {419}},\ \bibinfo {pages} {151}
		(\bibinfo {year} {2002})}\BibitemShut {NoStop}%
	\bibitem [{\citenamefont {Pruppacher}\ and\ \citenamefont
		{Klett}(2012)}]{pruppacher2012microphysics}%
	\BibitemOpen
	\bibfield  {author} {\bibinfo {author} {\bibfnamefont {H.~R.}\ \bibnamefont
			{Pruppacher}}\ and\ \bibinfo {author} {\bibfnamefont {J.~D.}\ \bibnamefont
			{Klett}},\ }\href@noop {} {\emph {\bibinfo {title} {Microphysics of clouds
				and precipitation: Reprinted 1980}}}\ (\bibinfo  {publisher} {Springer
		Science \& Business Media},\ \bibinfo {year} {2012})\BibitemShut {NoStop}%
	\bibitem [{\citenamefont {Williams}(1988)}]{williams1988unified}%
	\BibitemOpen
	\bibfield  {author} {\bibinfo {author} {\bibfnamefont {M.}~\bibnamefont
			{Williams}},\ }\bibfield  {title} {\bibinfo {title} {A unified theory of
			aerosol coagulation},\ }\href@noop {} {\bibfield  {journal} {\bibinfo
			{journal} {Journal of Physics D: Applied Physics}\ }\textbf {\bibinfo
			{volume} {21}},\ \bibinfo {pages} {875} (\bibinfo {year} {1988})}\BibitemShut
	{NoStop}%
	\bibitem [{\citenamefont {Brilliantov}\ \emph {et~al.}(2015)\citenamefont
		{Brilliantov}, \citenamefont {Krapivsky}, \citenamefont {Bodrova},
		\citenamefont {Spahn}, \citenamefont {Hayakawa}, \citenamefont {Stadnichuk},\
		and\ \citenamefont {Schmidt}}]{brilliantov2015size}%
	\BibitemOpen
	\bibfield  {author} {\bibinfo {author} {\bibfnamefont {N.}~\bibnamefont
			{Brilliantov}}, \bibinfo {author} {\bibfnamefont {P.}~\bibnamefont
			{Krapivsky}}, \bibinfo {author} {\bibfnamefont {A.}~\bibnamefont {Bodrova}},
		\bibinfo {author} {\bibfnamefont {F.}~\bibnamefont {Spahn}}, \bibinfo
		{author} {\bibfnamefont {H.}~\bibnamefont {Hayakawa}}, \bibinfo {author}
		{\bibfnamefont {V.}~\bibnamefont {Stadnichuk}},\ and\ \bibinfo {author}
		{\bibfnamefont {J.}~\bibnamefont {Schmidt}},\ }\bibfield  {title} {\bibinfo
		{title} {Size distribution of particles in saturn’s rings from aggregation
			and fragmentation},\ }\href@noop {} {\bibfield  {journal} {\bibinfo
			{journal} {Proceedings of the National Academy of Sciences}\ }\textbf
		{\bibinfo {volume} {112}},\ \bibinfo {pages} {9536} (\bibinfo {year}
		{2015})}\BibitemShut {NoStop}%
	\bibitem [{\citenamefont {Connaughton}\ \emph {et~al.}(2018)\citenamefont
		{Connaughton}, \citenamefont {Dutta}, \citenamefont {Rajesh}, \citenamefont
		{Siddharth},\ and\ \citenamefont {Zaboronski}}]{connaughton2018stationary}%
	\BibitemOpen
	\bibfield  {author} {\bibinfo {author} {\bibfnamefont {C.}~\bibnamefont
			{Connaughton}}, \bibinfo {author} {\bibfnamefont {A.}~\bibnamefont {Dutta}},
		\bibinfo {author} {\bibfnamefont {R.}~\bibnamefont {Rajesh}}, \bibinfo
		{author} {\bibfnamefont {N.}~\bibnamefont {Siddharth}},\ and\ \bibinfo
		{author} {\bibfnamefont {O.}~\bibnamefont {Zaboronski}},\ }\bibfield  {title}
	{\bibinfo {title} {Stationary mass distribution and nonlocality in models of
			coalescence and shattering},\ }\href@noop {} {\bibfield  {journal} {\bibinfo
			{journal} {Physical Review E}\ }\textbf {\bibinfo {volume} {97}},\ \bibinfo
		{pages} {022137} (\bibinfo {year} {2018})}\BibitemShut {NoStop}%
	\bibitem [{\citenamefont {Burd}\ and\ \citenamefont
		{Jackson}(2009)}]{burd2009particle}%
	\BibitemOpen
	\bibfield  {author} {\bibinfo {author} {\bibfnamefont {A.~B.}\ \bibnamefont
			{Burd}}\ and\ \bibinfo {author} {\bibfnamefont {G.~A.}\ \bibnamefont
			{Jackson}},\ }\bibfield  {title} {\bibinfo {title} {Particle aggregation},\
	}\href@noop {} {\bibfield  {journal} {\bibinfo  {journal} {Annual review of
				marine science}\ }\textbf {\bibinfo {volume} {1}},\ \bibinfo {pages} {65}
		(\bibinfo {year} {2009})}\BibitemShut {NoStop}%
	\bibitem [{\citenamefont {Benjwal}\ \emph {et~al.}(2006)\citenamefont
		{Benjwal}, \citenamefont {Verma}, \citenamefont {R{\"o}hm},\ and\
		\citenamefont {Gursky}}]{benjwal2006monitoring}%
	\BibitemOpen
	\bibfield  {author} {\bibinfo {author} {\bibfnamefont {S.}~\bibnamefont
			{Benjwal}}, \bibinfo {author} {\bibfnamefont {S.}~\bibnamefont {Verma}},
		\bibinfo {author} {\bibfnamefont {K.-H.}\ \bibnamefont {R{\"o}hm}},\ and\
		\bibinfo {author} {\bibfnamefont {O.}~\bibnamefont {Gursky}},\ }\bibfield
	{title} {\bibinfo {title} {Monitoring protein aggregation during thermal
			unfolding in circular dichroism experiments},\ }\href@noop {} {\bibfield
		{journal} {\bibinfo  {journal} {Protein Science}\ }\textbf {\bibinfo {volume}
			{15}},\ \bibinfo {pages} {635} (\bibinfo {year} {2006})}\BibitemShut
	{NoStop}%
	\bibitem [{\citenamefont {Wang}\ \emph {et~al.}(2015)\citenamefont {Wang},
		\citenamefont {Guo}, \citenamefont {Lou},\ and\ \citenamefont
		{Xu}}]{wang2015following}%
	\BibitemOpen
	\bibfield  {author} {\bibinfo {author} {\bibfnamefont {B.}~\bibnamefont
			{Wang}}, \bibinfo {author} {\bibfnamefont {C.}~\bibnamefont {Guo}}, \bibinfo
		{author} {\bibfnamefont {Z.}~\bibnamefont {Lou}},\ and\ \bibinfo {author}
		{\bibfnamefont {B.}~\bibnamefont {Xu}},\ }\bibfield  {title} {\bibinfo
		{title} {Following the aggregation of human prion protein on au (111) surface
			in real-time},\ }\href@noop {} {\bibfield  {journal} {\bibinfo  {journal}
			{Chemical Communications}\ }\textbf {\bibinfo {volume} {51}},\ \bibinfo
		{pages} {2088} (\bibinfo {year} {2015})}\BibitemShut {NoStop}%
	\bibitem [{\citenamefont {Tom}\ \emph {et~al.}(2016)\citenamefont {Tom},
		\citenamefont {Rajesh},\ and\ \citenamefont
		{Vemparala}}]{tom2016aggregation}%
	\BibitemOpen
	\bibfield  {author} {\bibinfo {author} {\bibfnamefont {A.~M.}\ \bibnamefont
			{Tom}}, \bibinfo {author} {\bibfnamefont {R.}~\bibnamefont {Rajesh}},\ and\
		\bibinfo {author} {\bibfnamefont {S.}~\bibnamefont {Vemparala}},\ }\bibfield
	{title} {\bibinfo {title} {Aggregation dynamics of rigid polyelectrolytes},\
	}\href@noop {} {\bibfield  {journal} {\bibinfo  {journal} {The Journal of
				Chemical Physics}\ }\textbf {\bibinfo {volume} {144}} (\bibinfo {year}
		{2016})}\BibitemShut {NoStop}%
	\bibitem [{\citenamefont {Tom}\ \emph {et~al.}(2017)\citenamefont {Tom},
		\citenamefont {Rajesh},\ and\ \citenamefont
		{Vemparala}}]{tom2017aggregation}%
	\BibitemOpen
	\bibfield  {author} {\bibinfo {author} {\bibfnamefont {A.~M.}\ \bibnamefont
			{Tom}}, \bibinfo {author} {\bibfnamefont {R.}~\bibnamefont {Rajesh}},\ and\
		\bibinfo {author} {\bibfnamefont {S.}~\bibnamefont {Vemparala}},\ }\bibfield
	{title} {\bibinfo {title} {Aggregation of flexible polyelectrolytes: Phase
			diagram and dynamics},\ }\href@noop {} {\bibfield  {journal} {\bibinfo
			{journal} {The Journal of chemical physics}\ }\textbf {\bibinfo {volume}
			{147}} (\bibinfo {year} {2017})}\BibitemShut {NoStop}%
	\bibitem [{\citenamefont {Pineau}\ \emph {et~al.}(2016)\citenamefont {Pineau},
		\citenamefont {Benzerga},\ and\ \citenamefont {Pardoen}}]{pineau2016failure}%
	\BibitemOpen
	\bibfield  {author} {\bibinfo {author} {\bibfnamefont {A.}~\bibnamefont
			{Pineau}}, \bibinfo {author} {\bibfnamefont {A.~A.}\ \bibnamefont
			{Benzerga}},\ and\ \bibinfo {author} {\bibfnamefont {T.}~\bibnamefont
			{Pardoen}},\ }\bibfield  {title} {\bibinfo {title} {Failure of metals i:
			Brittle and ductile fracture},\ }\href@noop {} {\bibfield  {journal}
		{\bibinfo  {journal} {Acta Materialia}\ }\textbf {\bibinfo {volume} {107}},\
		\bibinfo {pages} {424} (\bibinfo {year} {2016})}\BibitemShut {NoStop}%
	\bibitem [{\citenamefont {Smoluchowski}(1917)}]{smoluchowski1917mathematical}%
	\BibitemOpen
	\bibfield  {author} {\bibinfo {author} {\bibfnamefont {M.}~\bibnamefont
			{Smoluchowski}},\ }\bibfield  {title} {\bibinfo {title} {Mathematical theory
			of the kinetics of the coagulation of colloidal solutions},\ }\href@noop {}
	{\bibfield  {journal} {\bibinfo  {journal} {Z. Phys. Chem.}\ }\textbf
		{\bibinfo {volume} {92}},\ \bibinfo {pages} {129} (\bibinfo {year}
		{1917})}\BibitemShut {NoStop}%
	\bibitem [{\citenamefont {Tarboton}\ \emph {et~al.}(1988)\citenamefont
		{Tarboton}, \citenamefont {Bras},\ and\ \citenamefont
		{Rodriguez-Iturbe}}]{tarboton1988fractal}%
	\BibitemOpen
	\bibfield  {author} {\bibinfo {author} {\bibfnamefont {D.~G.}\ \bibnamefont
			{Tarboton}}, \bibinfo {author} {\bibfnamefont {R.~L.}\ \bibnamefont {Bras}},\
		and\ \bibinfo {author} {\bibfnamefont {I.}~\bibnamefont {Rodriguez-Iturbe}},\
	}\bibfield  {title} {\bibinfo {title} {The fractal nature of river
			networks},\ }\href@noop {} {\bibfield  {journal} {\bibinfo  {journal} {Water
				resources research}\ }\textbf {\bibinfo {volume} {24}},\ \bibinfo {pages}
		{1317} (\bibinfo {year} {1988})}\BibitemShut {NoStop}%
	\bibitem [{\citenamefont {Berestycki}(2009)}]{berestycki2009recent}%
	\BibitemOpen
	\bibfield  {author} {\bibinfo {author} {\bibfnamefont {N.}~\bibnamefont
			{Berestycki}},\ }\bibfield  {title} {\bibinfo {title} {Recent progress in
			coalescent theory},\ }\href@noop {} {\bibfield  {journal} {\bibinfo
			{journal} {ENSAIOS MATEM{\'A}TICOS}\ }\textbf {\bibinfo {volume} {16}},\
		\bibinfo {pages} {1} (\bibinfo {year} {2009})}\BibitemShut {NoStop}%
	\bibitem [{\citenamefont {Achlioptas}\ \emph {et~al.}(2009)\citenamefont
		{Achlioptas}, \citenamefont {D'souza},\ and\ \citenamefont
		{Spencer}}]{achlioptas2009explosive}%
	\BibitemOpen
	\bibfield  {author} {\bibinfo {author} {\bibfnamefont {D.}~\bibnamefont
			{Achlioptas}}, \bibinfo {author} {\bibfnamefont {R.~M.}\ \bibnamefont
			{D'souza}},\ and\ \bibinfo {author} {\bibfnamefont {J.}~\bibnamefont
			{Spencer}},\ }\bibfield  {title} {\bibinfo {title} {Explosive percolation in
			random networks},\ }\href@noop {} {\bibfield  {journal} {\bibinfo  {journal}
			{science}\ }\textbf {\bibinfo {volume} {323}},\ \bibinfo {pages} {1453}
		(\bibinfo {year} {2009})}\BibitemShut {NoStop}%
	\bibitem [{\citenamefont {D'Souza}\ \emph {et~al.}(2019)\citenamefont
		{D'Souza}, \citenamefont {G{\'o}mez-Gardenes}, \citenamefont {Nagler},\ and\
		\citenamefont {Arenas}}]{d2019explosive}%
	\BibitemOpen
	\bibfield  {author} {\bibinfo {author} {\bibfnamefont {R.~M.}\ \bibnamefont
			{D'Souza}}, \bibinfo {author} {\bibfnamefont {J.}~\bibnamefont
			{G{\'o}mez-Gardenes}}, \bibinfo {author} {\bibfnamefont {J.}~\bibnamefont
			{Nagler}},\ and\ \bibinfo {author} {\bibfnamefont {A.}~\bibnamefont
			{Arenas}},\ }\bibfield  {title} {\bibinfo {title} {Explosive phenomena in
			complex networks},\ }\href@noop {} {\bibfield  {journal} {\bibinfo  {journal}
			{Advances in Physics}\ }\textbf {\bibinfo {volume} {68}},\ \bibinfo {pages}
		{123} (\bibinfo {year} {2019})}\BibitemShut {NoStop}%
	\bibitem [{\citenamefont {Leyvraz}(2003)}]{leyvraz2003scaling}%
	\BibitemOpen
	\bibfield  {author} {\bibinfo {author} {\bibfnamefont {F.}~\bibnamefont
			{Leyvraz}},\ }\bibfield  {title} {\bibinfo {title} {Scaling theory and
			exactly solved models in the kinetics of irreversible aggregation},\
	}\href@noop {} {\bibfield  {journal} {\bibinfo  {journal} {Physics Reports}\
		}\textbf {\bibinfo {volume} {383}},\ \bibinfo {pages} {95} (\bibinfo {year}
		{2003})}\BibitemShut {NoStop}%
	\bibitem [{\citenamefont {Aldous}(1999)}]{aldous1999deterministic}%
	\BibitemOpen
	\bibfield  {author} {\bibinfo {author} {\bibfnamefont {D.~J.}\ \bibnamefont
			{Aldous}},\ }\bibfield  {title} {\bibinfo {title} {Deterministic and
			stochastic models for coalescence (aggregation and coagulation): a review of
			the mean-field theory for probabilists},\ }\href@noop {} {\bibfield
		{journal} {\bibinfo  {journal} {Bernoulli}\ }\textbf {\bibinfo {volume}
			{5}},\ \bibinfo {pages} {3} (\bibinfo {year} {1999})}\BibitemShut {NoStop}%
	\bibitem [{\citenamefont {Krapivsky}\ \emph {et~al.}(2010)\citenamefont
		{Krapivsky}, \citenamefont {Redner},\ and\ \citenamefont
		{Ben-Naim}}]{krapivsky2010kinetic}%
	\BibitemOpen
	\bibfield  {author} {\bibinfo {author} {\bibfnamefont {P.~L.}\ \bibnamefont
			{Krapivsky}}, \bibinfo {author} {\bibfnamefont {S.}~\bibnamefont {Redner}},\
		and\ \bibinfo {author} {\bibfnamefont {E.}~\bibnamefont {Ben-Naim}},\
	}\href@noop {} {\emph {\bibinfo {title} {A kinetic view of statistical
				physics}}}\ (\bibinfo  {publisher} {Cambridge University Press},\ \bibinfo
	{year} {2010})\BibitemShut {NoStop}%
	\bibitem [{\citenamefont {Wattis}(2006)}]{WATTIS20061}%
	\BibitemOpen
	\bibfield  {author} {\bibinfo {author} {\bibfnamefont {J.~A.}\ \bibnamefont
			{Wattis}},\ }\bibfield  {title} {\bibinfo {title} {An introduction to
			mathematical models of coagulation–fragmentation processes: A discrete
			deterministic mean-field approach},\ }\href
	{https://doi.org/https://doi.org/10.1016/j.physd.2006.07.024} {\bibfield
		{journal} {\bibinfo  {journal} {Physica D: Nonlinear Phenomena}\ }\textbf
		{\bibinfo {volume} {222}},\ \bibinfo {pages} {1} (\bibinfo {year} {2006})},\
	\bibinfo {note} {coagulation-fragmentation Processes}\BibitemShut {NoStop}%
	\bibitem [{\citenamefont {Connaughton}\ \emph {et~al.}(2010)\citenamefont
		{Connaughton}, \citenamefont {Rajesh},\ and\ \citenamefont
		{Zaboronski}}]{handbook}%
	\BibitemOpen
	\bibfield  {author} {\bibinfo {author} {\bibfnamefont {C.}~\bibnamefont
			{Connaughton}}, \bibinfo {author} {\bibfnamefont {R.}~\bibnamefont
			{Rajesh}},\ and\ \bibinfo {author} {\bibfnamefont {O.}~\bibnamefont
			{Zaboronski}},\ }\bibfield  {title} {\bibinfo {title} {Kinetics of
			cluster-cluster aggregation},\ }in\ \href
	{https://doi.org/https://doi.org/10.1201/9781420075557} {\emph {\bibinfo
			{booktitle} {Handbook of Nanophysics: Clusters and Fullerenes}}},\ \bibinfo
	{editor} {edited by\ \bibinfo {editor} {\bibfnamefont {K.~D.}\ \bibnamefont
			{Sattler}}}\ (\bibinfo  {publisher} {CRC Press},\ \bibinfo {year}
	{2010})\BibitemShut {NoStop}%
	\bibitem [{\citenamefont {Puthalath}\ \emph {et~al.}(2023)\citenamefont
		{Puthalath}, \citenamefont {Biswas}, \citenamefont {Prasad},\ and\
		\citenamefont {Rajesh}}]{puthalath2023lattice}%
	\BibitemOpen
	\bibfield  {author} {\bibinfo {author} {\bibfnamefont {F.}~\bibnamefont
			{Puthalath}}, \bibinfo {author} {\bibfnamefont {A.}~\bibnamefont {Biswas}},
		\bibinfo {author} {\bibfnamefont {V.}~\bibnamefont {Prasad}},\ and\ \bibinfo
		{author} {\bibfnamefont {R.}~\bibnamefont {Rajesh}},\ }\bibfield  {title}
	{\bibinfo {title} {Lattice models for ballistic aggregation:
			Cluster-shape-dependent exponents},\ }\href@noop {} {\bibfield  {journal}
		{\bibinfo  {journal} {Physical Review E}\ }\textbf {\bibinfo {volume}
			{108}},\ \bibinfo {pages} {044127} (\bibinfo {year} {2023})}\BibitemShut
	{NoStop}%
	\bibitem [{\citenamefont {Rajesh}\ \emph {et~al.}(2024)\citenamefont {Rajesh},
		\citenamefont {Subashri},\ and\ \citenamefont
		{Zaboronski}}]{rajesh2024exact}%
	\BibitemOpen
	\bibfield  {author} {\bibinfo {author} {\bibfnamefont {R.}~\bibnamefont
			{Rajesh}}, \bibinfo {author} {\bibfnamefont {V.}~\bibnamefont {Subashri}},\
		and\ \bibinfo {author} {\bibfnamefont {O.}~\bibnamefont {Zaboronski}},\
	}\bibfield  {title} {\bibinfo {title} {Exact calculation of the probabilities
			of rare events in cluster-cluster aggregation},\ }\href@noop {} {\bibfield
		{journal} {\bibinfo  {journal} {Physical Review Letters}\ }\textbf {\bibinfo
			{volume} {133}},\ \bibinfo {pages} {097101} (\bibinfo {year}
		{2024})}\BibitemShut {NoStop}%
	\bibitem [{\citenamefont {Dandekar}\ \emph {et~al.}(2023)\citenamefont
		{Dandekar}, \citenamefont {Rajesh}, \citenamefont {Subashri},\ and\
		\citenamefont {Zaboronski}}]{dandekar2023monte}%
	\BibitemOpen
	\bibfield  {author} {\bibinfo {author} {\bibfnamefont {R.}~\bibnamefont
			{Dandekar}}, \bibinfo {author} {\bibfnamefont {R.}~\bibnamefont {Rajesh}},
		\bibinfo {author} {\bibfnamefont {V.}~\bibnamefont {Subashri}},\ and\
		\bibinfo {author} {\bibfnamefont {O.}~\bibnamefont {Zaboronski}},\ }\bibfield
	{title} {\bibinfo {title} {A monte carlo algorithm to measure probabilities
			of rare events in cluster-cluster aggregation},\ }\href@noop {} {\bibfield
		{journal} {\bibinfo  {journal} {Computer Physics Communications}\ }\textbf
		{\bibinfo {volume} {288}},\ \bibinfo {pages} {108727} (\bibinfo {year}
		{2023})}\BibitemShut {NoStop}%
	\bibitem [{\citenamefont {Rajesh}\ \emph {et~al.}(2025)\citenamefont {Rajesh},
		\citenamefont {Subashri},\ and\ \citenamefont
		{Zaboronski}}]{rajesh2025exact}%
	\BibitemOpen
	\bibfield  {author} {\bibinfo {author} {\bibfnamefont {R.}~\bibnamefont
			{Rajesh}}, \bibinfo {author} {\bibfnamefont {V.}~\bibnamefont {Subashri}},\
		and\ \bibinfo {author} {\bibfnamefont {O.}~\bibnamefont {Zaboronski}},\
	}\bibfield  {title} {\bibinfo {title} {Exact calculation of the large
			deviation function for k-nary coalescence},\ }\href@noop {} {\bibfield
		{journal} {\bibinfo  {journal} {Journal of Statistical Physics}\ }\textbf
		{\bibinfo {volume} {192}},\ \bibinfo {pages} {72} (\bibinfo {year}
		{2025})}\BibitemShut {NoStop}%
	\bibitem [{\citenamefont {Andreis}\ \emph {et~al.}(2021)\citenamefont
		{Andreis}, \citenamefont {K{\"o}nig},\ and\ \citenamefont
		{Patterson}}]{andreis2021large}%
	\BibitemOpen
	\bibfield  {author} {\bibinfo {author} {\bibfnamefont {L.}~\bibnamefont
			{Andreis}}, \bibinfo {author} {\bibfnamefont {W.}~\bibnamefont {K{\"o}nig}},\
		and\ \bibinfo {author} {\bibfnamefont {R.~I.}\ \bibnamefont {Patterson}},\
	}\bibfield  {title} {\bibinfo {title} {A large-deviations principle for all
			the cluster sizes of a sparse erd{\H{o}}s--r{\'e}nyi graph},\ }\href@noop {}
	{\bibfield  {journal} {\bibinfo  {journal} {Random Structures \& Algorithms}\
		}\textbf {\bibinfo {volume} {59}},\ \bibinfo {pages} {522} (\bibinfo {year}
		{2021})}\BibitemShut {NoStop}%
	\bibitem [{\citenamefont {Andreis}\ \emph {et~al.}(2023)\citenamefont
		{Andreis}, \citenamefont {K{\"o}nig}, \citenamefont {Langhammer},\ and\
		\citenamefont {Patterson}}]{andreis2023large}%
	\BibitemOpen
	\bibfield  {author} {\bibinfo {author} {\bibfnamefont {L.}~\bibnamefont
			{Andreis}}, \bibinfo {author} {\bibfnamefont {W.}~\bibnamefont {K{\"o}nig}},
		\bibinfo {author} {\bibfnamefont {H.}~\bibnamefont {Langhammer}},\ and\
		\bibinfo {author} {\bibfnamefont {R.~I.}\ \bibnamefont {Patterson}},\
	}\bibfield  {title} {\bibinfo {title} {A large-deviations principle for all
			the components in a sparse inhomogeneous random graph},\ }\href@noop {}
	{\bibfield  {journal} {\bibinfo  {journal} {Probability Theory and Related
				Fields}\ }\textbf {\bibinfo {volume} {186}},\ \bibinfo {pages} {521}
		(\bibinfo {year} {2023})}\BibitemShut {NoStop}%
	\bibitem [{\citenamefont {Engel}\ \emph {et~al.}(2004)\citenamefont {Engel},
		\citenamefont {Monasson},\ and\ \citenamefont {Hartmann}}]{engel2004large}%
	\BibitemOpen
	\bibfield  {author} {\bibinfo {author} {\bibfnamefont {A.}~\bibnamefont
			{Engel}}, \bibinfo {author} {\bibfnamefont {R.}~\bibnamefont {Monasson}},\
		and\ \bibinfo {author} {\bibfnamefont {A.~K.}\ \bibnamefont {Hartmann}},\
	}\bibfield  {title} {\bibinfo {title} {On large deviation properties of
			erd{\"o}s--r{\'e}nyi random graphs},\ }\href@noop {} {\bibfield  {journal}
		{\bibinfo  {journal} {Journal of Statistical Physics}\ }\textbf {\bibinfo
			{volume} {117}},\ \bibinfo {pages} {387} (\bibinfo {year}
		{2004})}\BibitemShut {NoStop}%
	\bibitem [{\citenamefont {Marcus}(1968)}]{marcus1968stochastic}%
	\BibitemOpen
	\bibfield  {author} {\bibinfo {author} {\bibfnamefont {A.~H.}\ \bibnamefont
			{Marcus}},\ }\bibfield  {title} {\bibinfo {title} {Stochastic coalescence},\
	}\href@noop {} {\bibfield  {journal} {\bibinfo  {journal} {Technometrics}\
		}\textbf {\bibinfo {volume} {10}},\ \bibinfo {pages} {133} (\bibinfo {year}
		{1968})}\BibitemShut {NoStop}%
	\bibitem [{\citenamefont {Lushnikov}(2006)}]{lushnikov2006gelation}%
	\BibitemOpen
	\bibfield  {author} {\bibinfo {author} {\bibfnamefont {A.~A.}\ \bibnamefont
			{Lushnikov}},\ }\bibfield  {title} {\bibinfo {title} {Gelation in coagulating
			systems},\ }\href@noop {} {\bibfield  {journal} {\bibinfo  {journal} {Physica
				D: Nonlinear Phenomena}\ }\textbf {\bibinfo {volume} {222}},\ \bibinfo
		{pages} {37} (\bibinfo {year} {2006})}\BibitemShut {NoStop}%
	\bibitem [{\citenamefont {Lushnikov}(1973)}]{lushnikov1973evolution}%
	\BibitemOpen
	\bibfield  {author} {\bibinfo {author} {\bibfnamefont {A.~A.}\ \bibnamefont
			{Lushnikov}},\ }\bibfield  {title} {\bibinfo {title} {Evolution of
			coagulating systems},\ }\href@noop {} {\bibfield  {journal} {\bibinfo
			{journal} {Journal of Colloid and Interface Science}\ }\textbf {\bibinfo
			{volume} {45}},\ \bibinfo {pages} {549} (\bibinfo {year} {1973})}\BibitemShut
	{NoStop}%
	\bibitem [{\citenamefont {Lushnikov}(1978)}]{lushnikov1978coagulation}%
	\BibitemOpen
	\bibfield  {author} {\bibinfo {author} {\bibfnamefont {A.~A.}\ \bibnamefont
			{Lushnikov}},\ }\bibfield  {title} {\bibinfo {title} {Coagulation in finite
			systems},\ }\href@noop {} {\bibfield  {journal} {\bibinfo  {journal} {Journal
				of Colloid and interface science}\ }\textbf {\bibinfo {volume} {65}},\
		\bibinfo {pages} {276} (\bibinfo {year} {1978})}\BibitemShut {NoStop}%
	\bibitem [{\citenamefont {van Dongen}\ and\ \citenamefont
		{Ernst}(1985)}]{van1985dynamic}%
	\BibitemOpen
	\bibfield  {author} {\bibinfo {author} {\bibfnamefont {P.~G.}\ \bibnamefont
			{van Dongen}}\ and\ \bibinfo {author} {\bibfnamefont {M.~H.}\ \bibnamefont
			{Ernst}},\ }\bibfield  {title} {\bibinfo {title} {Dynamic scaling in the
			kinetics of clustering},\ }\href@noop {} {\bibfield  {journal} {\bibinfo
			{journal} {Physical review letters}\ }\textbf {\bibinfo {volume} {54}},\
		\bibinfo {pages} {1396} (\bibinfo {year} {1985})}\BibitemShut {NoStop}%
	\bibitem [{\citenamefont {Lushnikov}(2005)}]{lushnikov2005exact}%
	\BibitemOpen
	\bibfield  {author} {\bibinfo {author} {\bibfnamefont {A.~A.}\ \bibnamefont
			{Lushnikov}},\ }\bibfield  {title} {\bibinfo {title} {Exact kinetics of the
			sol-gel transition},\ }\href@noop {} {\bibfield  {journal} {\bibinfo
			{journal} {Physical Review E—Statistical, Nonlinear, and Soft Matter
				Physics}\ }\textbf {\bibinfo {volume} {71}},\ \bibinfo {pages} {046129}
		(\bibinfo {year} {2005})}\BibitemShut {NoStop}%
	\bibitem [{\citenamefont {Ball}\ \emph {et~al.}(2012)\citenamefont {Ball},
		\citenamefont {Connaughton}, \citenamefont {Jones}, \citenamefont {Rajesh},\
		and\ \citenamefont {Zaboronski}}]{ball2012collective}%
	\BibitemOpen
	\bibfield  {author} {\bibinfo {author} {\bibfnamefont {R.~C.}\ \bibnamefont
			{Ball}}, \bibinfo {author} {\bibfnamefont {C.}~\bibnamefont {Connaughton}},
		\bibinfo {author} {\bibfnamefont {P.~P.}\ \bibnamefont {Jones}}, \bibinfo
		{author} {\bibfnamefont {R.}~\bibnamefont {Rajesh}},\ and\ \bibinfo {author}
		{\bibfnamefont {O.}~\bibnamefont {Zaboronski}},\ }\bibfield  {title}
	{\bibinfo {title} {Collective oscillations in irreversible coagulation driven
			by monomer inputs and large-cluster outputs},\ }\href@noop {} {\bibfield
		{journal} {\bibinfo  {journal} {Physical review letters}\ }\textbf {\bibinfo
			{volume} {109}},\ \bibinfo {pages} {168304} (\bibinfo {year}
		{2012})}\BibitemShut {NoStop}%
	\bibitem [{\citenamefont {Dedola}\ and\ \citenamefont
		{Cademartiri}(2026)}]{dedola2026gelation}%
	\BibitemOpen
	\bibfield  {author} {\bibinfo {author} {\bibfnamefont {M.}~\bibnamefont
			{Dedola}}\ and\ \bibinfo {author} {\bibfnamefont {L.}~\bibnamefont
			{Cademartiri}},\ }\bibfield  {title} {\bibinfo {title} {Is gelation a
			singularity or a flow induced instability?},\ }\href@noop {} {\bibfield
		{journal} {\bibinfo  {journal} {arXiv preprint arXiv:2601.18806}\ } (\bibinfo
		{year} {2026})}\BibitemShut {NoStop}%
	\bibitem [{\citenamefont {Alsmeyer}\ \emph {et~al.}(2023)\citenamefont
		{Alsmeyer}, \citenamefont {Bostan}, \citenamefont {Raschel},\ and\
		\citenamefont {Simon}}]{alsmeyer2023persistence}%
	\BibitemOpen
	\bibfield  {author} {\bibinfo {author} {\bibfnamefont {G.}~\bibnamefont
			{Alsmeyer}}, \bibinfo {author} {\bibfnamefont {A.}~\bibnamefont {Bostan}},
		\bibinfo {author} {\bibfnamefont {K.}~\bibnamefont {Raschel}},\ and\ \bibinfo
		{author} {\bibfnamefont {T.}~\bibnamefont {Simon}},\ }\bibfield  {title}
	{\bibinfo {title} {Persistence for a class of order-one autoregressive
			processes and mallows-riordan polynomials},\ }\href@noop {} {\bibfield
		{journal} {\bibinfo  {journal} {Advances in Applied Mathematics}\ }\textbf
		{\bibinfo {volume} {150}},\ \bibinfo {pages} {102555} (\bibinfo {year}
		{2023})}\BibitemShut {NoStop}%
	\bibitem [{\citenamefont {Touchette}(2009)}]{touchette2009large}%
	\BibitemOpen
	\bibfield  {author} {\bibinfo {author} {\bibfnamefont {H.}~\bibnamefont
			{Touchette}},\ }\bibfield  {title} {\bibinfo {title} {The large deviation
			approach to statistical mechanics},\ }\href@noop {} {\bibfield  {journal}
		{\bibinfo  {journal} {Physics Reports}\ }\textbf {\bibinfo {volume} {478}},\
		\bibinfo {pages} {1} (\bibinfo {year} {2009})}\BibitemShut {NoStop}%
	\bibitem [{\citenamefont {Griffiths}\ and\ \citenamefont
		{Gujrati}(1983)}]{griffiths1983convexity}%
	\BibitemOpen
	\bibfield  {author} {\bibinfo {author} {\bibfnamefont {R.~B.}\ \bibnamefont
			{Griffiths}}\ and\ \bibinfo {author} {\bibfnamefont {P.}~\bibnamefont
			{Gujrati}},\ }\bibfield  {title} {\bibinfo {title} {Convexity violations for
			noninteger parameters in certain lattice models},\ }\href@noop {} {\bibfield
		{journal} {\bibinfo  {journal} {Journal of Statistical Physics}\ }\textbf
		{\bibinfo {volume} {30}},\ \bibinfo {pages} {563} (\bibinfo {year}
		{1983})}\BibitemShut {NoStop}%
	\bibitem [{\citenamefont {Gartner}(1977)}]{gartner1977large}%
	\BibitemOpen
	\bibfield  {author} {\bibinfo {author} {\bibfnamefont {J.}~\bibnamefont
			{Gartner}},\ }\bibfield  {title} {\bibinfo {title} {On large deviations from
			the invariant measure},\ }\href@noop {} {\bibfield  {journal} {\bibinfo
			{journal} {Theory of Probability and its Applications}\ }\textbf {\bibinfo
			{volume} {22}},\ \bibinfo {pages} {24} (\bibinfo {year} {1977})}\BibitemShut
	{NoStop}%
	\bibitem [{\citenamefont {Ellis}(1984)}]{ellis1984large}%
	\BibitemOpen
	\bibfield  {author} {\bibinfo {author} {\bibfnamefont {R.~S.}\ \bibnamefont
			{Ellis}},\ }\bibfield  {title} {\bibinfo {title} {Large deviations for a
			general class of random vectors},\ }\href@noop {} {\bibfield  {journal}
		{\bibinfo  {journal} {The Annals of Probability}\ }\textbf {\bibinfo {volume}
			{12}},\ \bibinfo {pages} {1} (\bibinfo {year} {1984})}\BibitemShut {NoStop}%
	\bibitem [{\citenamefont {Rockafellar}(2015)}]{rockafellar2015convex}%
	\BibitemOpen
	\bibfield  {author} {\bibinfo {author} {\bibfnamefont {R.~T.}\ \bibnamefont
			{Rockafellar}},\ }\href@noop {} {\emph {\bibinfo {title} {Convex Analysis}}}\
	(\bibinfo  {publisher} {Princeton University Press},\ \bibinfo {address}
	{Princeton, NJ},\ \bibinfo {year} {2015})\BibitemShut {NoStop}%
	\bibitem [{\citenamefont {Krapivsky}\ and\ \citenamefont
		{Matveev}(2024)}]{krapivsky2024gelation}%
	\BibitemOpen
	\bibfield  {author} {\bibinfo {author} {\bibfnamefont {P.~L.}\ \bibnamefont
			{Krapivsky}}\ and\ \bibinfo {author} {\bibfnamefont {S.~A.}\ \bibnamefont
			{Matveev}},\ }\bibfield  {title} {\bibinfo {title} {Gelation in input-driven
			aggregation},\ }\href@noop {} {\bibfield  {journal} {\bibinfo  {journal}
			{Physical Review E}\ }\textbf {\bibinfo {volume} {110}},\ \bibinfo {pages}
		{034128} (\bibinfo {year} {2024})}\BibitemShut {NoStop}%
	\bibitem [{\citenamefont {Dyachenko}\ \emph {et~al.}(2023)\citenamefont
		{Dyachenko}, \citenamefont {Matveev},\ and\ \citenamefont
		{Krapivsky}}]{dyachenko2023finite}%
	\BibitemOpen
	\bibfield  {author} {\bibinfo {author} {\bibfnamefont {R.~R.}\ \bibnamefont
			{Dyachenko}}, \bibinfo {author} {\bibfnamefont {S.~A.}\ \bibnamefont
			{Matveev}},\ and\ \bibinfo {author} {\bibfnamefont {P.~L.}\ \bibnamefont
			{Krapivsky}},\ }\bibfield  {title} {\bibinfo {title} {Finite-size effects in
			addition and chipping processes},\ }\href@noop {} {\bibfield  {journal}
		{\bibinfo  {journal} {Physical Review E}\ }\textbf {\bibinfo {volume}
			{108}},\ \bibinfo {pages} {044119} (\bibinfo {year} {2023})}\BibitemShut
	{NoStop}%
	\bibitem [{\citenamefont {Spouge}(1988)}]{spouge1988exact}%
	\BibitemOpen
	\bibfield  {author} {\bibinfo {author} {\bibfnamefont {J.~L.}\ \bibnamefont
			{Spouge}},\ }\bibfield  {title} {\bibinfo {title} {Exact solutions for a
			diffusion-reaction process in one dimension},\ }\href@noop {} {\bibfield
		{journal} {\bibinfo  {journal} {Physical review letters}\ }\textbf {\bibinfo
			{volume} {60}},\ \bibinfo {pages} {871} (\bibinfo {year} {1988})}\BibitemShut
	{NoStop}%
	\bibitem [{\citenamefont {Kang}\ and\ \citenamefont
		{Redner}(1984)}]{kang1984fluctuation}%
	\BibitemOpen
	\bibfield  {author} {\bibinfo {author} {\bibfnamefont {K.}~\bibnamefont
			{Kang}}\ and\ \bibinfo {author} {\bibfnamefont {S.}~\bibnamefont {Redner}},\
	}\bibfield  {title} {\bibinfo {title} {Fluctuation effects in smoluchowski
			reaction kinetics},\ }\href@noop {} {\bibfield  {journal} {\bibinfo
			{journal} {Physical Review A}\ }\textbf {\bibinfo {volume} {30}},\ \bibinfo
		{pages} {2833} (\bibinfo {year} {1984})}\BibitemShut {NoStop}%
	\bibitem [{\citenamefont {Krishnamurthy}\ \emph {et~al.}(2002)\citenamefont
		{Krishnamurthy}, \citenamefont {Rajesh},\ and\ \citenamefont
		{Zaboronski}}]{krishnamurthy2002kang}%
	\BibitemOpen
	\bibfield  {author} {\bibinfo {author} {\bibfnamefont {S.}~\bibnamefont
			{Krishnamurthy}}, \bibinfo {author} {\bibfnamefont {R.}~\bibnamefont
			{Rajesh}},\ and\ \bibinfo {author} {\bibfnamefont {O.}~\bibnamefont
			{Zaboronski}},\ }\bibfield  {title} {\bibinfo {title} {Kang-redner small-mass
			anomaly in cluster-cluster aggregation},\ }\href@noop {} {\bibfield
		{journal} {\bibinfo  {journal} {Physical Review E}\ }\textbf {\bibinfo
			{volume} {66}},\ \bibinfo {pages} {066118} (\bibinfo {year}
		{2002})}\BibitemShut {NoStop}%
	\bibitem [{\citenamefont {Krishnamurthy}\ \emph {et~al.}(2003)\citenamefont
		{Krishnamurthy}, \citenamefont {Rajesh},\ and\ \citenamefont
		{Zaboronski}}]{krishnamurthy2003persistence}%
	\BibitemOpen
	\bibfield  {author} {\bibinfo {author} {\bibfnamefont {S.}~\bibnamefont
			{Krishnamurthy}}, \bibinfo {author} {\bibfnamefont {R.}~\bibnamefont
			{Rajesh}},\ and\ \bibinfo {author} {\bibfnamefont {O.}~\bibnamefont
			{Zaboronski}},\ }\bibfield  {title} {\bibinfo {title} {Persistence properties
			of a system of coagulating and annihilating random walkers},\ }\href@noop {}
	{\bibfield  {journal} {\bibinfo  {journal} {Physical Review E}\ }\textbf
		{\bibinfo {volume} {68}},\ \bibinfo {pages} {046103} (\bibinfo {year}
		{2003})}\BibitemShut {NoStop}%
	\bibitem [{\citenamefont {Ravindran}\ and\ \citenamefont
		{Rajesh}(2025)}]{ravindran2025finite}%
	\BibitemOpen
	\bibfield  {author} {\bibinfo {author} {\bibfnamefont {R.~B.}\ \bibnamefont
			{Ravindran}}\ and\ \bibinfo {author} {\bibfnamefont {R.}~\bibnamefont
			{Rajesh}},\ }\bibfield  {title} {\bibinfo {title} {Finite size scaling and
			edge effects in the takayasu model of aggregation diffusion with input},\
	}\href@noop {} {\bibfield  {journal} {\bibinfo  {journal} {arXiv preprint
				arXiv:2511.09470}\ } (\bibinfo {year} {2025})}\BibitemShut {NoStop}%
\end{thebibliography}
%

\end{document}